\documentclass[iop]{emulateapj}

\usepackage{natbib,aas_macros,amsmath}
\usepackage{multirow,color}
\usepackage{natbib}

\newcommand{\carcsec}{$\!\!\arcsec$}
\newcommand{\m}[1]{\mathrm{#1}}

\newcommand{\redc}[1]{\textcolor{black}{#1}}

\begin{document}

\shortauthors{Harikane et al.}
\slugcomment{Accepted for Publication in The Astrophysical Journal}

\shorttitle{
Large Scale Structures with Protoclusters at $z\sim 6-7$
}

\title{
SILVERRUSH. VIII. Spectroscopic Identifications of Early Large Scale Structures with Protoclusters Over 200 M\lowercase{pc} 
at $\lowercase{z}\sim 6-7$: Strong Associations of Dusty Star-Forming Galaxies
}

\email{hari@icrr.u-tokyo.ac.jp}
\author{
Yuichi Harikane\altaffilmark{1,2,3},
Masami Ouchi\altaffilmark{1,4},
Yoshiaki Ono\altaffilmark{1},
Seiji Fujimoto\altaffilmark{1,5},
Darko Donevski\altaffilmark{6,7},
Takatoshi Shibuya\altaffilmark{8},
Andreas L. Faisst\altaffilmark{9},
Tomotsugu Goto\altaffilmark{10},
Bunyo Hatsukade\altaffilmark{11},
Nobunari Kashikawa\altaffilmark{5},
Kotaro Kohno\altaffilmark{11},
Takuya Hashimoto\altaffilmark{12,3},
Ryo Higuchi\altaffilmark{1,2},
Akio K. Inoue\altaffilmark{12},
Yen-Ting Lin\altaffilmark{13},
Crystal L. Martin\altaffilmark{14},
Roderik Overzier\altaffilmark{15,16},
Ian Smail\altaffilmark{17},
Jun Toshikawa\altaffilmark{1},
Hideki Umehata\altaffilmark{18,11},
Yiping Ao\altaffilmark{19},
Scott Chapman\altaffilmark{20},
David L. Clements\altaffilmark{21},
Myungshin Im\altaffilmark{22},
Yipeng Jing\altaffilmark{23,24},
Toshihiro Kawaguchi\altaffilmark{25},
Chien-Hsiu Lee\altaffilmark{26},
Minju M. Lee\altaffilmark{27,3},
Lihwai Lin\altaffilmark{13},
Yoshiki Matsuoka\altaffilmark{28},
Murilo Marinello\altaffilmark{15},
Tohru Nagao\altaffilmark{29},
Masato Onodera\altaffilmark{26},
Sune Toft\altaffilmark{29},
Wei-Hao Wang\altaffilmark{13}
}

\affil{$^1$
Institute for Cosmic Ray Research, The University of Tokyo, 5-1-5 Kashiwanoha, Kashiwa, Chiba 277-8582, Japan
}
\affil{$^2$
Department of Physics, Graduate School of Science, The University of Tokyo, 7-3-1 Hongo, Bunkyo, Tokyo, 113-0033, Japan
}
\affil{$^3$
National Astronomical Observatory of Japan, 2-21-1 Osawa, Mitaka, Tokyo 181-8588, Japan
}
\affil{$^4$
Kavli Institute for the Physics and Mathematics of the Universe (WPI), University of Tokyo, Kashiwa 277-8583, Japan
}
\affil{$^5$
Department of Astronomy, Graduate School of Science, The University of Tokyo, 7-3-1 Hongo, Bunkyo, Tokyo 113-0033, Japan
}
\affil{$^6$
Aix Marseille University, CNRS, LAM, Laboratoire d’Astrophysique de Marseille, Marseille, France
}
\affil{$^7$
SISSA, via Bonomea 265, I-34136 Trieste, Italy
}
\affil{$^8$
Kitami Institute of Technology, 165 Koen-cho, Kitami, Hokkaido 090-8507, Japan
}
\affil{$^9$
Infrared Processing and Analysis Center, California Institute of Technology, MC 100-22, 770 South Wilson Ave., Pasadena, CA 91125, USA
}
\affil{$^{10}$
Institute of Astronomy, National Tsing Hua University, No. 101, Section 2, Kuang-Fu Road, Hsinchu, Taiwan
}
\affil{$^{11}$
Institute of Astronomy, Graduate School of Science, The University of Tokyo, 2-21-1 Osawa, Mitaka, Tokyo 181-0015, Japan
}
\affil{$^{12}$
Department of Environmental Science and Technology, Faculty of Design Technology, Osaka Sangyo University, 3-1-1, Nagaito, Daito, Osaka 574-8530, Japan
}
\affil{$^{13}$
Institute of Astronomy \& Astrophysics, Academia Sinica, Taipei 106, Taiwan (ROC)
}
\affil{$^{14}$
Department of Physics, University of California, Santa Barbara, CA, 93106, USA
}
\affil{$^{15}$
Observatorio Nacional, Rua Jose Cristino, 77. CEP 20921-400, Sao Cristovao, Rio de Janeiro-RJ, Brazil
}
\affil{$^{16}$
Universidade de Sao Paulo, Instituto de Astronomia, Geof\'{\i}sica e Ci\^{e}ncias Atmosf\'{e}ricas, Departamento de Astronomia, S\~{a}o
Paulo, SP 05508-090, Brazil
}
\affil{$^{17}$
Centre for Extragalactic Astronomy, Department of Physics, Durham University, South Road, Durham DH1 3LE, UK
}
\affil{$^{18}$
RIKEN Cluster for Pioneering Research, 2-1 Hirosawa, Wako-shi, Saitama 351-0198, Japan
}
\affil{$^{19}$
Purple Mountain Observatory \& Key Laboratory for Radio Astronomy, Chinese Academy of Sciences, 8 Yuanhua Road, Nanjing 210034, China
}
\affil{$^{20}$
Department of Physics and Atmospheric Science, Dalhousie University, Halifax, NS B3H 3J5 Canada
}
\affil{$^{21}$
Astrophysics Group, Imperial College London, Blackett Laboratory, Prince Consort Road, London SW7 2AZ, UK
}
\affil{$^{22}$
CEOU/Astronomy Program, Dept. of Physics \& Astronomy, Seoul National University, Seoul, Korea
}
\affil{$^{23}$
School of Physics and Astronomy, Tsung-Dao Lee Institute, Shanghai Jiao Tong University, 800 Dongchuan Road, Shanghai, 200240,China
}
\affil{$^{24}$
IFSA Collaborative Innovation Center, Shanghai Jiao Tong University, Shanghai 200240, China
}
\affil{$^{25}$
Department of Economics, Management and Information Science, Onomichi City University, Hisayamada 1600-2, Onomichi, Hiroshima 722-8506, Japan
}
\affil{$^{26}$
Subaru Telescope, NAOJ, 650 N Aohoku Pl, Hilo, HI 96720, USA
}
\affil{$^{27}$
Division of Particle and Astrophysical Science, Graduate School of Science, Nagoya University, Furo-cho, Chikusa-ku, Nagoya 464-8602, Japan
}
\affil{$^{28}$
Research Center for Space and Cosmic Evolution, Ehime University, Bunkyo-cho, Matsuyama, Ehime 790-8577, Japan
}
\affil{$^{29}$
Cosmic Dawn Center (DAWN), Niels Bohr Institute, Juliane Mariesvej 30, DK-2100 Copenhagen, Denmark
}

\begin{abstract}
We have obtained three-dimensional maps of the universe in $\sim200\times200\times80$ comoving Mpc$^3$ (cMpc$^3$) volumes each at $z=5.7$ and $6.6$ based on a spectroscopic sample of 179 galaxies that achieves $\gtrsim80$\% completeness down to the Ly$\alpha$ luminosity of $\log(L_{\rm Ly\alpha}/[\mathrm{erg\ s^{-1}}])=43.0$, \redc{based on our Keck and Gemini observations and the literature}.
The maps reveal filamentary large-scale structures and two remarkable overdensities made out of at least 44 and 12 galaxies at $z=5.692$ (z57OD) and $z=6.585$ (z66OD), respectively, making z66OD the most distant overdensity spectroscopically confirmed to date \redc{with $>10$ spectroscopically confirmed galaxies}.
We compare spatial distributions of submillimeter galaxies at $z\simeq 4-6$ with our $z=5.7$ galaxies forming the large-scale structures, and detect a $99.97\%$ signal of cross correlation, indicative of a clear coincidence of dusty star-forming galaxy and dust unobscured galaxy formation at this early epoch.
The galaxies in z57OD and z66OD are actively forming stars with star formation rates (SFRs) $\gtrsim5$ times higher than the main sequence, and particularly the SFR density in z57OD is 10 times higher than the cosmic average at the redshift (a.k.a. the Madau-Lilly plot).
Comparisons with numerical simulations suggest that z57OD and z66OD are protoclusters that are progenitors of the present-day clusters with halo masses of $\sim10^{14}\ \mathrm{M_\odot}$.
\end{abstract}


\section{Introduction}\label{ss_intro}
Galaxies are not uniformly distributed in the universe.
Some of them reside in groups and clusters on scales of $\sim1-3\ \m{Mpc}$, while others lie in long filaments of galaxies extending over $10\ \m{Mpc}$, called large scale structure \citep[e.g.,][]{2005ApJ...624..463G}.
Investigating the large scale structure is important for understanding galaxy formation, since there is observational evidence that galaxy properties depend on their environment.
Indeed at low redshift, galaxies in clusters are mostly passive, early-type galaxies \citep[e.g.,][]{1980ApJ...236..351D,2003MNRAS.346..601G}, and there is a clear trend that the star formation activity of galaxies tends to be lower in high-density environment than low-density environment \citep{2002MNRAS.334..673L,2004AJ....128.2677T}, known as the morphology/star formation-density relation.
Since galaxies in dense environments appear to experience accelerated evolution, we need to go to higher redshifts to study the progenitors of low redshift high-density environments.

Indeed, studies of the large scale structure at high redshift has shown that galaxies in dense regions experience enhanced star formation \citep[e.g.,][]{2001ApJ...562L...9K,2007A&A...468...33E,2010ApJ...719L.126T,2013MNRAS.434..423K}, opposite to the relation at low redshift.
In addition, recent cosmological simulations predict significant increase of the contribution to the cosmic star formation density from galaxy overdensities \citep{2017ApJ...844L..23C}.
Thus, many galaxy overdensities have been identified and investigated at $z>1$ to date, including protoclusters that grow to cluster-scale halos at the present day \citep[e.g.,][see \citealt{2016A&ARv..24...14O} for a review]{1998ApJ...492..428S,2005ApJ...626...44S,2003ApJ...586L.111S,2004ApJ...605L..93S,2014ApJ...782L...3C,2015ApJ...808...37C,2016ApJ...823...11D}.
At $z>3$, since strong rest-frame optical emission lines are redshifted to mid-infrared, the Ly$\alpha$ emission line is used as a spectroscopic probe for galaxies.
Some of the high-redshift overdense regions are identified with UV continuum and/or Ly$\alpha$ emission lines \citep[e.g.,][]{2006ApJ...637...58O,2010ApJ...721.1680U,2016ApJ...826..114T,2018ApJ...861...43P,2018arXiv180100531H}, and spectroscopically confirmed with Ly$\alpha$ \citep[e.g.,][]{2002ApJ...569L..11V,2005ApJ...620L...1O,2016ApJ...823...11D,2018NatAs.tmp..140J}, including the galaxy overdensities at $z=6.01$ \citep{2012ApJ...750..137T,2014ApJ...792...15T}.

Since the Ly$\alpha$ photons are easily absorbed by dust, it is important to investigate whether dust-obscured galaxies are also residing in high-redshift overdensities traced with the Ly$\alpha$ emission.
In addition, dusty star forming galaxies, such as submillimeter galaxies (SMGs), are expected to trace the most massive dark matter halos and overdensities at $z>2$ \citep[e.g.,][]{2016ApJ...824...36C,2017A&A...607A..89B,2018Natur.556..469M}.
\citet{2009Natur.459...61T} report $2.2\sigma$ large scale correlation between SMGs and Ly$\alpha$ emitters (LAEs) at $z=3.1$ in the SSA22 protocluster.
\citet{2014MNRAS.440.3462U} improve the selection of SMGs using photometric redshifts, and detect stronger correlation between SMGs and LAEs in the SSA22 protocluster \citep[see also;][]{2015ApJ...815L...8U,2017ApJ...834L..16U,2018PASJ...70...65U}.
These results suggest that dust-obscured star forming galaxies are also lying in the SSA22 protocluster traced by LAEs at $z=3.1$.
However, the association between SMGs and LAEs at higher redshift is not yet understood.

In this study, we investigate large scale structures at $z=5.7$ and $6.6$ in the SXDS field using a large spectroscopic sample of 179 LAEs.
Combined with our recent Keck/DEIMOS and Gemini/GMOS observations, we produce 3D maps of the universe traced with the LAEs in two $\sim200\times200\times80\ \m{cMpc^3}$ volumes at $z=5.7$ and $6.6$.
We investigate the correlation between the LAEs and dust-obscured high-redshift SMGs, and stellar populations to probe the environmental dependence of galaxy properties.
We also compare our observational results with recent numerical simulations.
\redc{
One of the large scale structures investigated in this study is a protocluster at $z=5.7$ firstly reported in \citet{2005ApJ...620L...1O}.
\citet{2005ApJ...620L...1O} spectroscopically confirm 15 LAEs around this protocluster.
Recently, \citet{2018NatAs.tmp..140J} study this protocluster with 46 spectroscopically-confirmed LAEs in the SXDS field.
In this study, we use 135 LAEs spectroscopically confirmed at $z=5.7$, which allows us to obtain more complete view of the 3D structure of this protocluster.
In addition, we will investigate the correlation with high-redshift SMGs that are not investigated in these studies.
}
This study is one in a series of papers from a program studying high redshift galaxies, named Systematic Identification of LAEs for Visible Exploration and Reionization Research Using Subaru HSC \citep[SILVERRUSH][]{2018PASJ...70S..13O}.
Early results are already reported in several papers \citep{2018PASJ...70S..13O,2018PASJ...70S..14S,2018PASJ...70S..15S,2018PASJ...70S..16K,2018ApJ...859...84H,2018PASJ...70...55I,2018arXiv180100531H}.

This paper is organized as follows.
In Section \ref{ss_sample}, we present our LAE sample.
We describe our spectroscopic observations in Section \ref{ss_specobs}.
We present our results in Section \ref{ss_results}, and in Section \ref{ss_summary} we summarize our findings.
Throughout this paper we use the recent Planck cosmological parameter sets constrained with the temperature power spectrum, temperature-polarization cross spectrum, polarization power spectrum, low-$l$ polarization, CMB lensing, and external data \citep[TT, TE, EE+lowP+lensing+ext result; ][]{2016A&A...594A..13P}:
$\Omega_\m{m}=0.3089$, $\Omega_\Lambda=0.6911$, $\Omega_\m{b}=0.049$, $h=0.6774$, and $\sigma_8=0.8159$.
We assume a \citet{2003PASP..115..763C} initial mass function (IMF) with lower and
upper mass cutoffs of $0.1\m{M_\odot}$ and $100\m{M_\odot}$, respectively.
All magnitudes are in the AB system \citep{1983ApJ...266..713O}.

\begin{figure*}
\begin{center}
  \includegraphics[clip,bb=0 10 1000 500,width=1\hsize]{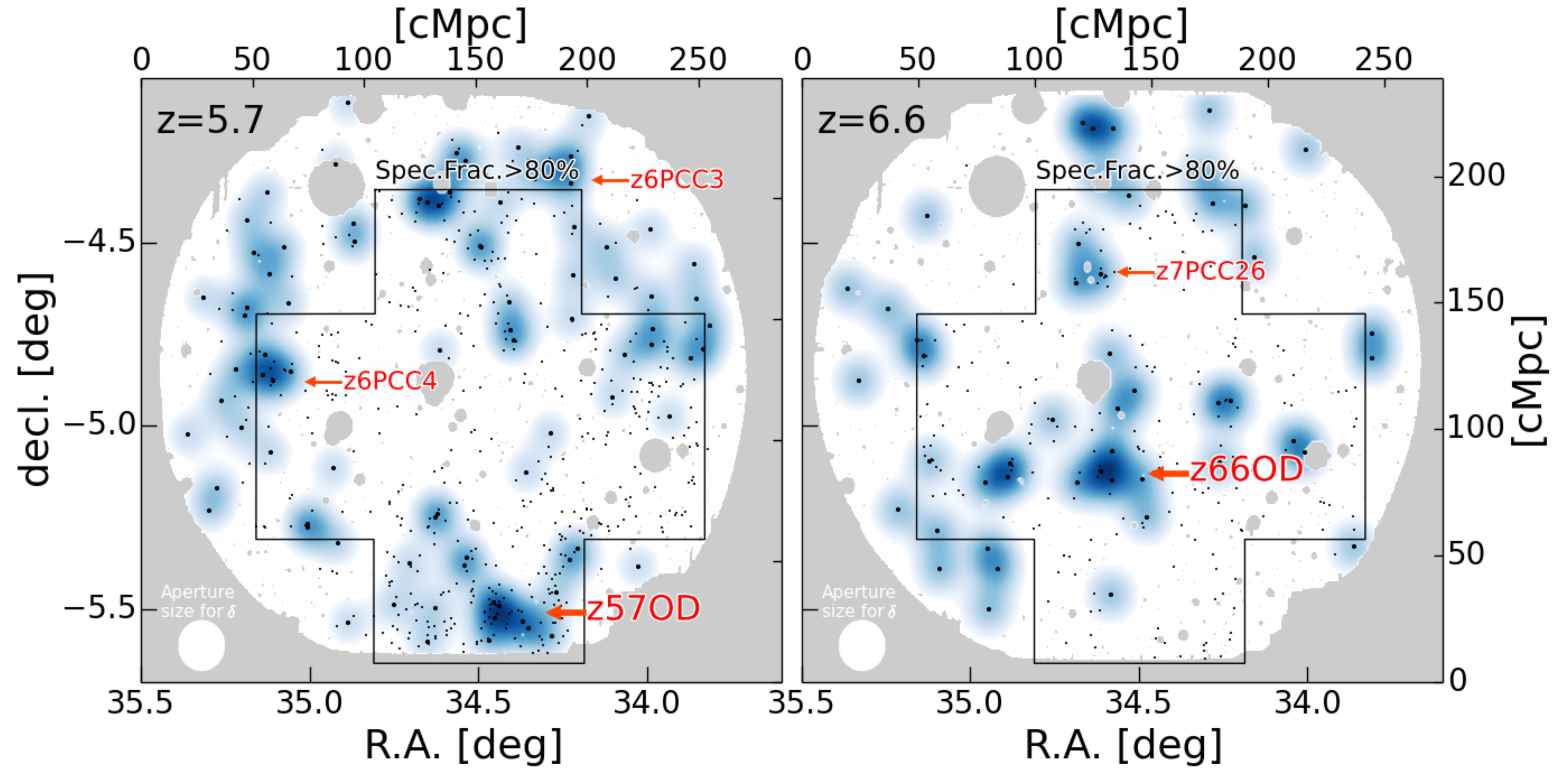}
 \end{center}
   \caption{Overdensity maps of LAEs at $z=5.7$ (left) and $z=6.6$ (right).
The black dots show the positions of the LAEs.
The large dots are LAEs whose NB magnitudes are brighter than 24.5 and 25.0 at $z=5.7$ and $6.6$, respectively.
The blue contours show number densities of LAEs brighter than 24.5 and 25.0 at $z=5.7$ and $6.6$, respectively.
Higher density regions are indicated by the darker colors.
The gray regions are masked due to the survey edges and bright stars.
The region indicated by the black polygon is the region where the fraction of spectroscopically confirmed LAEs brighter than $L_\mathrm{Ly\alpha}>10^{43}\ \m{erg\ s^{-1}}$ is $\gtrsim80\ \%$.
   \label{fig_map2d}}
\end{figure*}

\section{LAE Sample}\label{ss_sample}
We use LAE samples at $z=5.7$ and $6.6$ \citep{2018PASJ...70S..14S} selected based on the Subaru/Hyper Suprime-Cam Subaru strategic program (HSC-SSP) survey data \citep{2018PASJ...70S...8A,2018PASJ...70S...4A}, reduced with the HSC data processing pipeline \citep{2017arXiv170506766B}.
\redc{The LAEs at $z=5.7$ and $6.6$ are selected with the narrowband filters, NB816 and NB921, which have central wavelengths of $8170\ \m{\AA}$ and  $9210\ \m{\AA}$, and FWHMs of $131\ \m{\AA}$ and $120\ \m{\AA}$ to identify LAEs in the redshift range of $z=5.64-5.76$ and $z=6.50-6.63$, respectively.}
The HSC-SSP survey has three layers, UltraDeep (UD), Deep, and Wide, with different combinations of
area and depth.
In this study, we use LAE samples in the UD-SXDS field, where rich spectroscopic data are available (see Section \ref{ss_specobs}).
224 and 58 LAEs are selected at $z=5.7$ and $z=6.6$, respectively, in the UD-SXDS field with the following color criteria:\\
$z=5.7:$
\begin{eqnarray}
NB816<NB816_{5\sigma}\ \m{and}\ i-NB816>1.2\ \m{and}\notag\\
g>g_{3\sigma}\ \m{and}\ \left[(r\leq r_{3\sigma}\ \m{and}\  r-i\geq1.0)\ \m{or}\ r>r_{3\sigma}\right],\label{eq_colorz57}
\end{eqnarray}
$z=6.6:$
\begin{eqnarray}
NB921<NB921_{5\sigma}\ \m{and}\ z-NB921>1.0\ \m{and}\notag\\
g>g_{3\sigma}\ \m{and}\ r>r_{3\sigma}\ \m{and}\notag\\
\left[(z\leq z_{3\sigma}\ \m{and}\  i-z\geq1.3)\ \m{or}\ z>z_{3\sigma}\right].\label{eq_colorz66}
\end{eqnarray}
The subscripts ``$5\sigma$" and ``$3\sigma$" indicate the $5\sigma$ and $3\sigma$ magnitude limits for a given filter, respectively.
\redc{
Based on spectroscopic observations in \citet{2018PASJ...70S..15S}, the contamination rate is $0-30\%$.}
In addition, we use fainter LAE samples at $z=5.7$ and $6.6$ selected with Subaru/Suprime-Cam images in \citet{2008ApJS..176..301O,2010ApJ...723..869O}.
The total numbers of LAEs are 563 and 247 at $z=5.7$ and $6.6$, respectively.


To identify LAE overdensities, we calculate the galaxy overdensity, $\delta$, that is defined as follows:
\begin{equation}
\delta=\frac{n-\overline{n}}{\overline{n}},
\end{equation}
where $n$ is the number of LAEs in a cylinder and $\overline{n}$ is its average.
To draw two dimensional (2D) projected overdensity contours, we choose a cylinder whose height is $\sim40\ \m{cMpc}$ corresponding to the redshift range of the narrowband selected LAEs at each redshift.
The radius of the cylinder is $0.07\ \m{deg}$ which corresponds to $\sim10\ \m{cMpc}$ at $z\sim6$, which is a typical size of the protoclusters growing to $\sim10^{15}\ \m{M_\odot}$ halo at $z=0$ in simulations in \citet{2013ApJ...779..127C}.
We use LAEs brighter than $NB816<24.5$ and $NB921<25.0$ at $z=5.7$ and $6.6$, respectively, to keep high detection completeness.  
The average numbers of LAEs in a cylinder are $\overline{n}=0.48$ and $0.26$ at $z=5.7$ and $6.6$, respectively.
\redc{The masked regions are excluded in the calculations.}
In Figure \ref{fig_map2d}, we plot the calculated overdensities smoothed with a Gaussian kernel of $\sigma=0.07\ \m{deg}$.
Here we define overdensities as regions whose overdensity significances are higher than $4\sigma$ levels.
We identify overdensities previously reported in \citet{2018arXiv180100531H}; z6PCC1, z6PCC3, and z6PCC4 at $z=5.7$, and z7PCC24 and z7PCC26 (see also \citealt{2017MNRAS.469.2646C,2019ApJ...877...51C}) at $z=6.6$.\footnote{We regard z7PCC26 as an overdensity following \citet{2018arXiv180100531H}.}
z6PCC1 is the same structures reported in \citet{2005ApJ...620L...1O} and \citet[][see Section \ref{ss_3d}]{2018NatAs.tmp..140J}.
Hereafter we refer to z6PCC1 ($n=6$, $\delta=11.5$, $7.2\sigma$) and z7PCC24 ($n=4$, $\delta=14.3$, $6.8\sigma$), the most overdense regions at $z=5.7$ and $6.6$ in the UD-SXDS field, as z57OD and z66OD, respectively.

\section{Spectroscopic Data}\label{ss_specobs}

Out of 563 and 247 LAEs at $z=5.7$ and $6.6$, 135 and 36 LAEs are spectroscopically confirmed, respectively, in previous studies \citep{2005ApJ...620L...1O,2008ApJS..176..301O,2010ApJ...723..869O,2018PASJ...70S..15S,2018ApJ...859...84H,2018arXiv180100531H,2018NatAs.tmp..140J}.
Four LAEs around z66OD, z66LAE-1, -2, -3, and -4 are already spectroscopically confirmed.
In addition, we conducted Gemini and Keck spectroscopy targeting LAEs of z66OD.

We used Gemini Multi-Object Spectrographs (GMOS) on the $8 \m{m}$ Gemini North telescope in 2017 and 2018.
We used a total of two GMOS masks with the OG515 filter and R831 grating, and the total exposure times were 5400 and 10220 seconds.
Our exposures were conducted with spectral dithering of $50\ \m{\AA}$ to fill CCD gaps.
The spectroscopic coverage was between $7900\ \m{\AA}$ and $10000\ \m{\AA}$.
The spatial pixel scale was $0.\carcsec0727\ \m{pixel^{-1}}$.
The slit width was $0.\carcsec75$ and the spectral resolution was $R\sim3000$.
The seeing was around $0.\carcsec9$.
The reduction was performed using the Gemini {\sc iraf} packages\footnote{https://www.gemini.edu/sciops/data-and-results/processing-software}.
Wavelength calibration was achieved by fitting to the OH emission lines.

We also used DEep Imaging Multi-Object Spectrograph (DEIMOS) on the $10 \m{m}$ Keck II telescope in 2018.
We used one DEIMOS mask covering nine LAEs in z66OD, and the OG515 filter and R831 grating, and the total exposure times were 5400 and 10220 seconds.
We used one DEIMOS mask with the OG550 filter and 830G grating, and the total exposure time was 4900 seconds.
The spectroscopic coverage was between $6000\ \m{\AA}$ and $10000\ \m{\AA}$.
The spatial pixel scale was $0.\carcsec1185\ \m{pixel^{-1}}$.
The slit width was $0.\carcsec8$ and the spectral resolution was $R\sim3000$.
The seeing was around $0.\carcsec8$.
The reduction was performed using the {\tt spec2d} IDL pipeline developed by the DEEP2 Redshift Survey Team \citep{2003SPIE.4834..161D}.
Wavelength calibration was achieved by fitting to the arc lamp emission lines.

In these observations, we identified emission lines in eight LAEs, z66LAE-5, -6, -7, -8, -9, -10, -11 and -12.
We evaluate asymmetric profiles of these emission lines by calculating the weighted skewness, $S_w$ \citep{2006ApJ...648....7K}.
We find that the weighted skewness values of the lines in six LAEs, z66LAE-5, -6, -7, -8, -10, and -11 are larger than $3$, indicating that these asymmetric lines are Ly$\alpha$.
The weighted skewness values of the lines in z66LAE-9 and z66LAE-12 are less than $3$.
The narrow emission lines ($\m{FWHM}\simeq200\ \m{km\ s^{-1}}$ after a correction for the instrumental broadening) and medium spectral resolution ($R\sim3000$) do not suggest that these emission lines are {\sc [Oii]}$\lambda\lambda3726,3729$.
We do not find significant emission lines except for these lines at  $\sim9190$ and $\sim9250\ \m{\AA}$, rejecting the possibility of {[\sc Oiii]$\lambda$5007} emitters with detectable {[\sc Oiii]$\lambda$4959} or H$\beta$ lines, or H$\alpha$ emitters with a detectable {[\sc Oiii]$\lambda$5007} line.
Since most unresolved single line emitters have been found to be LAEs with a moderate velocity dispersion \citep{2004AJ....127..563H}, we regard these lines as Ly$\alpha$.
Note that removing z66LAE-9 and z66LAE-12 from our analysis does not change our conclusions.



\begin{figure}
 \begin{center}
  \includegraphics[clip,bb=0 0 300 440,width=1\hsize]{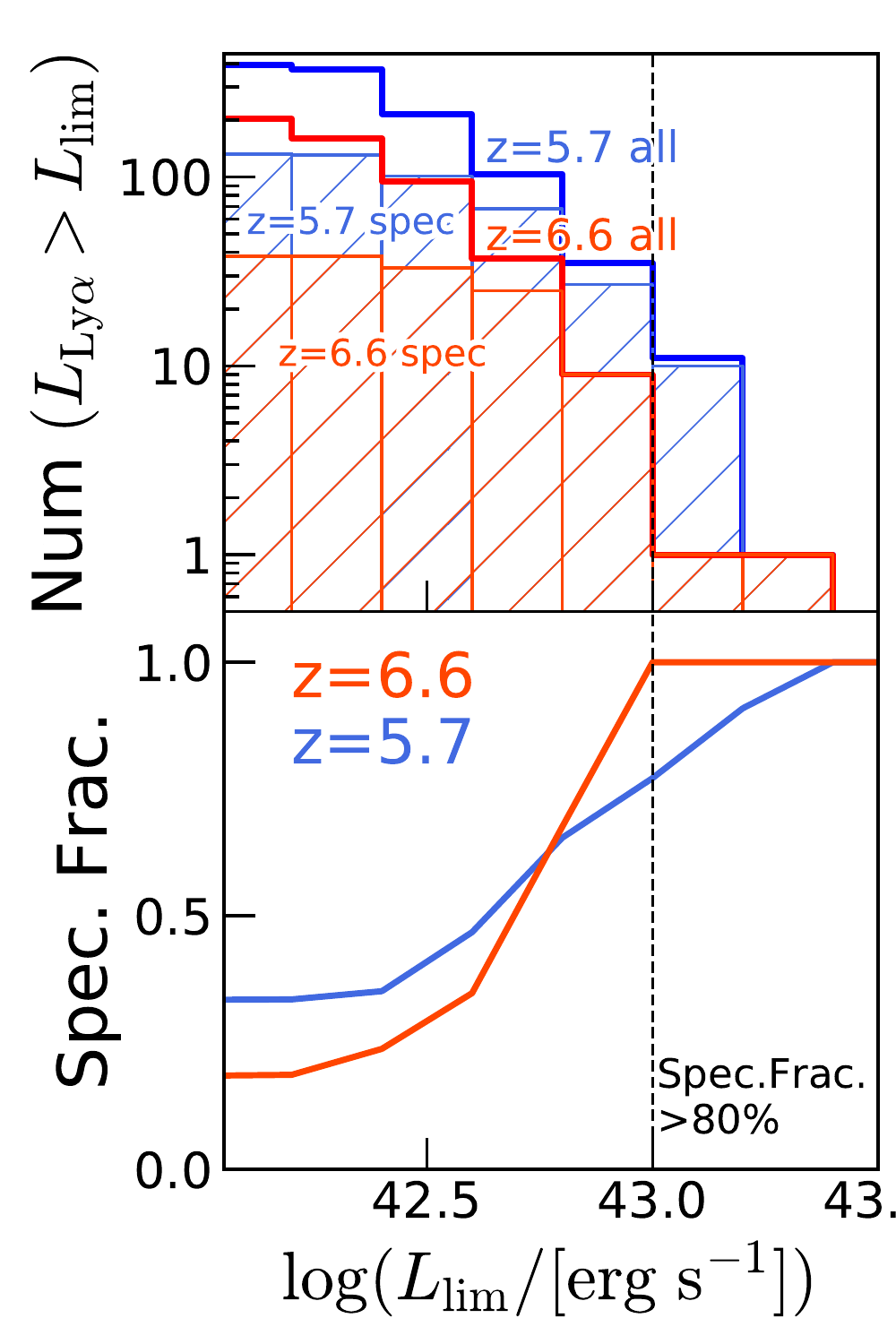}
 \end{center}
   \caption{
   {\bf Upper panel:}
  \redc{ Number of LAEs with spectroscopic confirmations.
   The blue and red histograms show cumulative numbers of all LAEs (open histogram) and spectroscopically confirmed LAEs (hatched histogram) in the black pentagon in Figure \ref{fig_map2d} at $z=5.7$ and $6.6$, respectively.}
   {\bf Lower panels:}
Fraction of spectroscopically confirmed LAEs at $z=5.7$ and $6.6$.
The blue and red solid curves show cumulative fractions of spectroscopically confirmed LAEs in the black pentagon in Figure \ref{fig_map2d} at $z=5.7$ and $6.6$, respectively.
   \label{fig_speccomp}}
\end{figure}

Thus total of 135 and 44 LAEs at $z=5.7$ and $6.6$ are used in this study.
\redc{Figure \ref{fig_speccomp} shows the numbers of LAEs spectroscopically confirmed and their fractions.}
Thanks to the large spectroscopic sample, the fraction of the \redc{spectroscopically confirmed LAEs} is $\gtrsim80\ \%$ down to the Ly$\alpha$ luminosity of $L_\mathrm{Ly\alpha}=10^{43}\ \m{erg\ s^{-1}}$ at $z=5.7$ and $6.6$ in the regions indicated with the black polygon in Figure \ref{fig_map2d}, corresponding to the SXDS fields in \cite{2008ApJS..176..301O,2010ApJ...723..869O}.
Although the \redc{spectroscopic fraction} of z57OD ($88\%$ for $L_\mathrm{Ly\alpha}>10^{43}\ \m{erg\ s^{-1}}$) is higher than that of all $z=5.7$ LAEs ($77\%$ for $L_\mathrm{Ly\alpha}>10^{43}\ \m{erg\ s^{-1}}$), the difference ($\sim10\%$) is not significant for our identifications of the overdensities in Section \ref{ss_3d}.
We do not find strong AGN signatures, such as the broad Ly$\alpha$ emission lines nor N{\sc v}1240 lines, in the spectra of our LAEs.

\section{Results and Discussions}\label{ss_results}

\begin{figure*}
\begin{center}
  \begin{minipage}{0.48\hsize}
 \begin{center}
  \includegraphics[clip,bb=0 60 340 260,width=1\hsize]{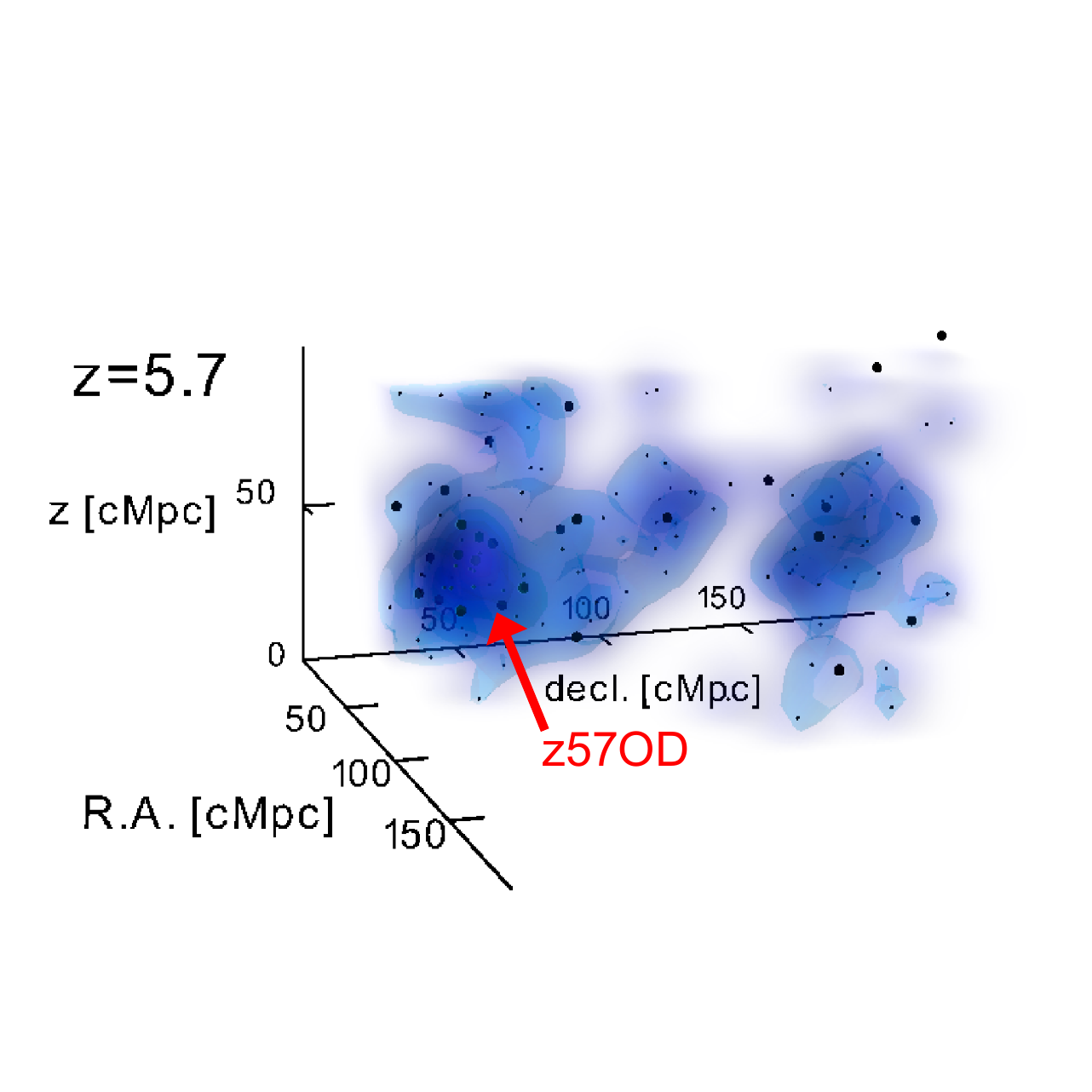}
 \end{center}
 \end{minipage}
 \begin{minipage}{0.48\hsize}
 \begin{center}
  \includegraphics[clip,bb=0 60 340 260,width=1\hsize]{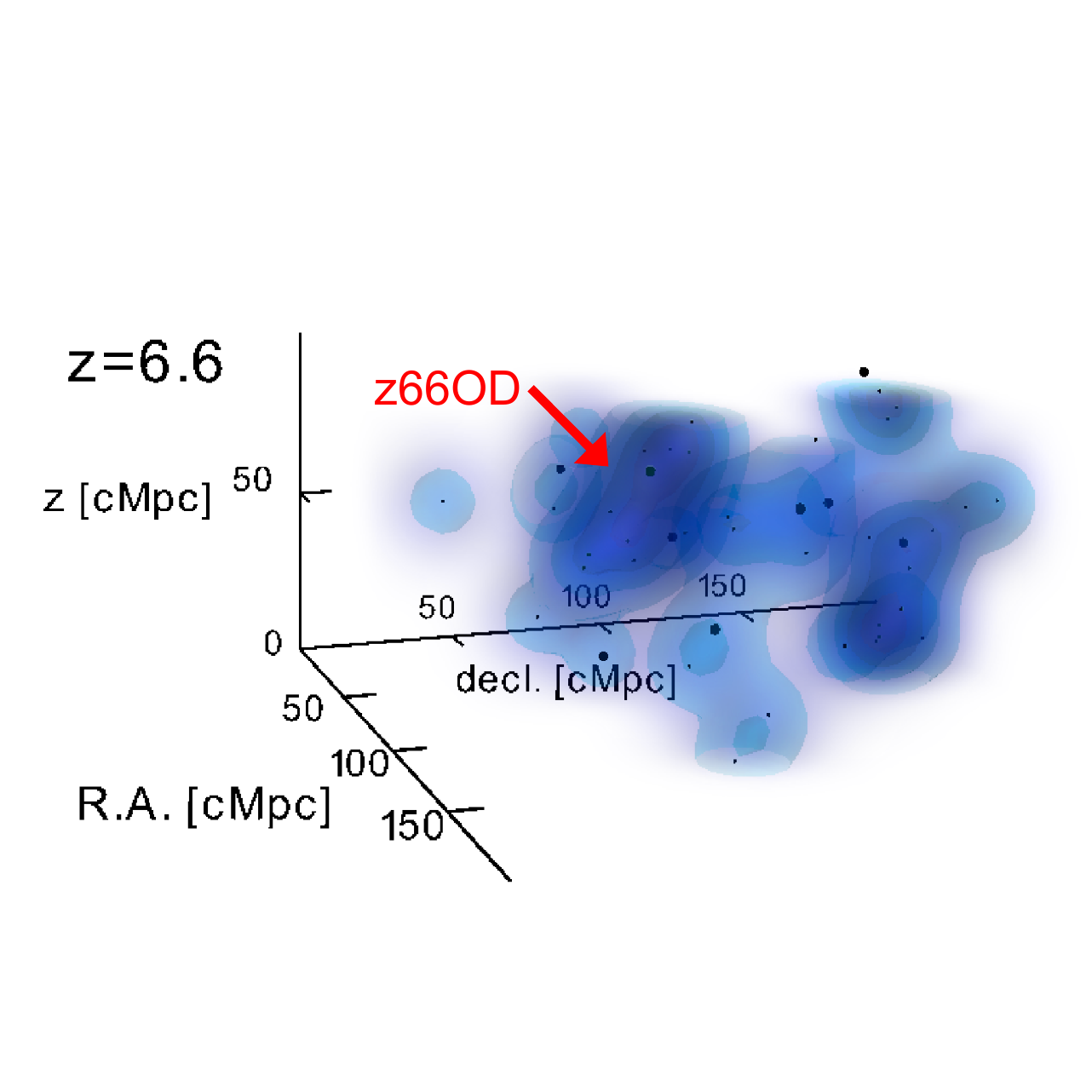}
 \end{center}
 \end{minipage}
 \end{center}
   \caption{3D overdensity maps of LAEs at $z=5.7$ (left) and $z=6.6$ (right).
   The black dots show the positions of the LAEs.
The large dots are LAEs brighter than $L_\mathrm{Ly\alpha}>10^{43}\ \m{erg\ s^{-1}}$.
Higher density regions are indicated by the bluer colors, smoothed with a Gaussian kernel of $\sigma=10\ \m{cMpc}$ ($15\ \m{cMpc}$) at $z=5.7$ ($z=6.6$).
   \label{fig_map3d}}
\end{figure*}

\begin{figure*}
\begin{center}
  \includegraphics[clip,bb=0 10 1900 800,width=1\hsize]{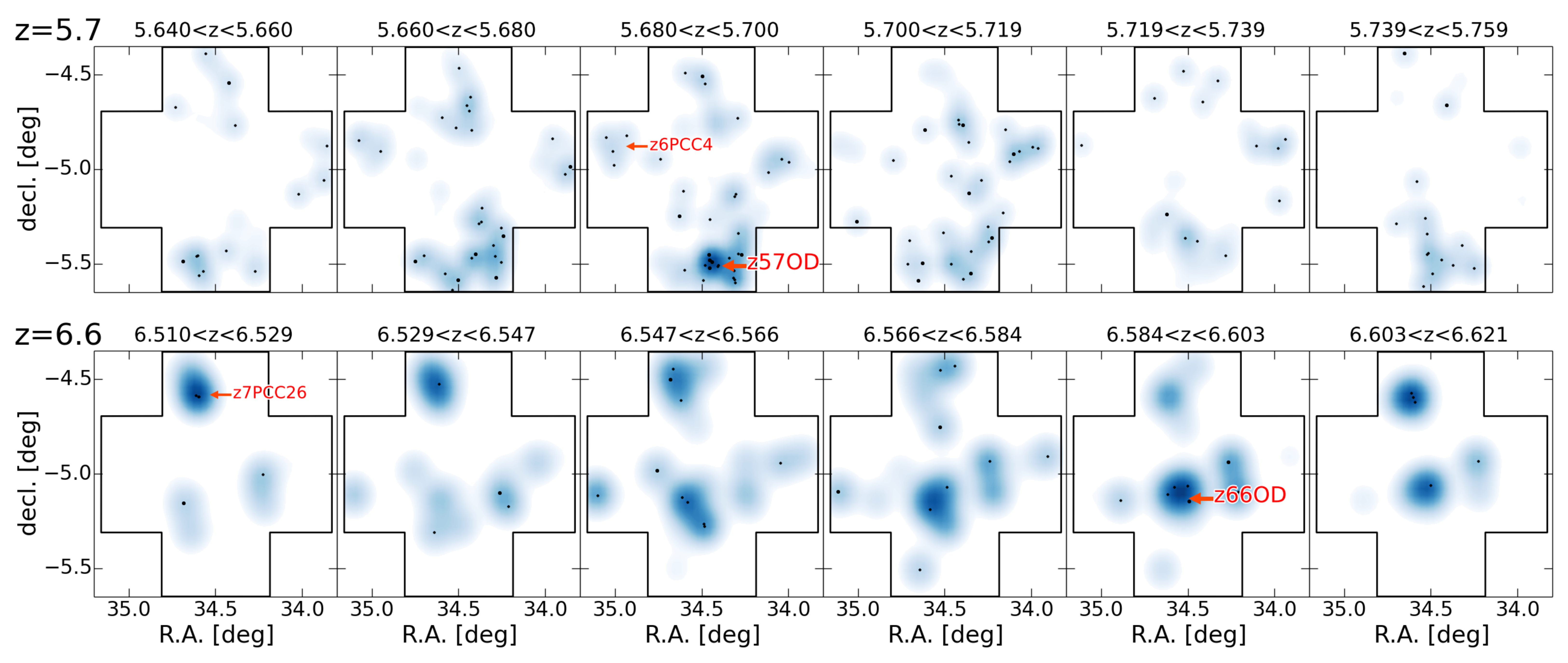}
 \end{center}
   \caption{
Two-dimensional map of LAEs at $z=5.7$ (upper) and $z=6.6$ (lower) with the redshift slices.
The black dots show the positions of the LAEs in the $\Delta z\sim0.02$ redshift depth.
The large dots are LAEs brighter than $L_\mathrm{Ly\alpha}>10^{43}\ \m{erg\ s^{-1}}$.
Higher density regions are indicated by the darker colors, smoothed with a Gaussian kernel of $\sigma=10\ \m{cMpc}$ ($15\ \m{cMpc}$) at $z=5.7$ ($z=6.6$).
   \label{fig_mapslice}}
\end{figure*}

\subsection{Large Scale Structure at $z=5.7$ and $z=6.6$ and Spectroscopic Confirmation of z66OD at $z=6.585$}\label{ss_3d}
We obtain the three-dimensional (3D) map using the 179 spectroscopic confirmed LAEs.
We calculate the 3D overdensity using the LAE sample with a sphere whose radius is $10\ \m{cMpc}$ ($15\ \m{cMpc}$) at $z=5.7$ ($z=6.6$).
Note that velocity offsets of the Ly$\alpha$ emission lines to the systemic redshifts are typically $\sim300\ \m{km\ s^{-1}}$ or $\sim2.5\ \m{cMpc}$ \citep[e.g.,][]{2014ApJ...795...33E,2016ApJ...822...29F,2018arXiv180600486H}, smaller than the radius of the sphere.
In Figure \ref{fig_map3d}, we plot the locations of the LAEs and the 3D overdensity smoothed with a Gaussian kernel of $\sigma=10\ \m{cMpc}$ ($15\ \m{cMpc}$) at $z=5.7$ ($z=6.6$).
Figure \ref{fig_mapslice} shows the 2D maps with the redshift slices of $\Delta z\sim0.02$.
These maps reveal the filamentary 3D large scale structures made by the LAEs at $z=5.7$ and $6.6$.

\begin{figure*}
 \begin{center}
  \begin{minipage}{0.41\hsize}
 \begin{center}
  \includegraphics[clip,bb=0 40 430 915,width=1\hsize]{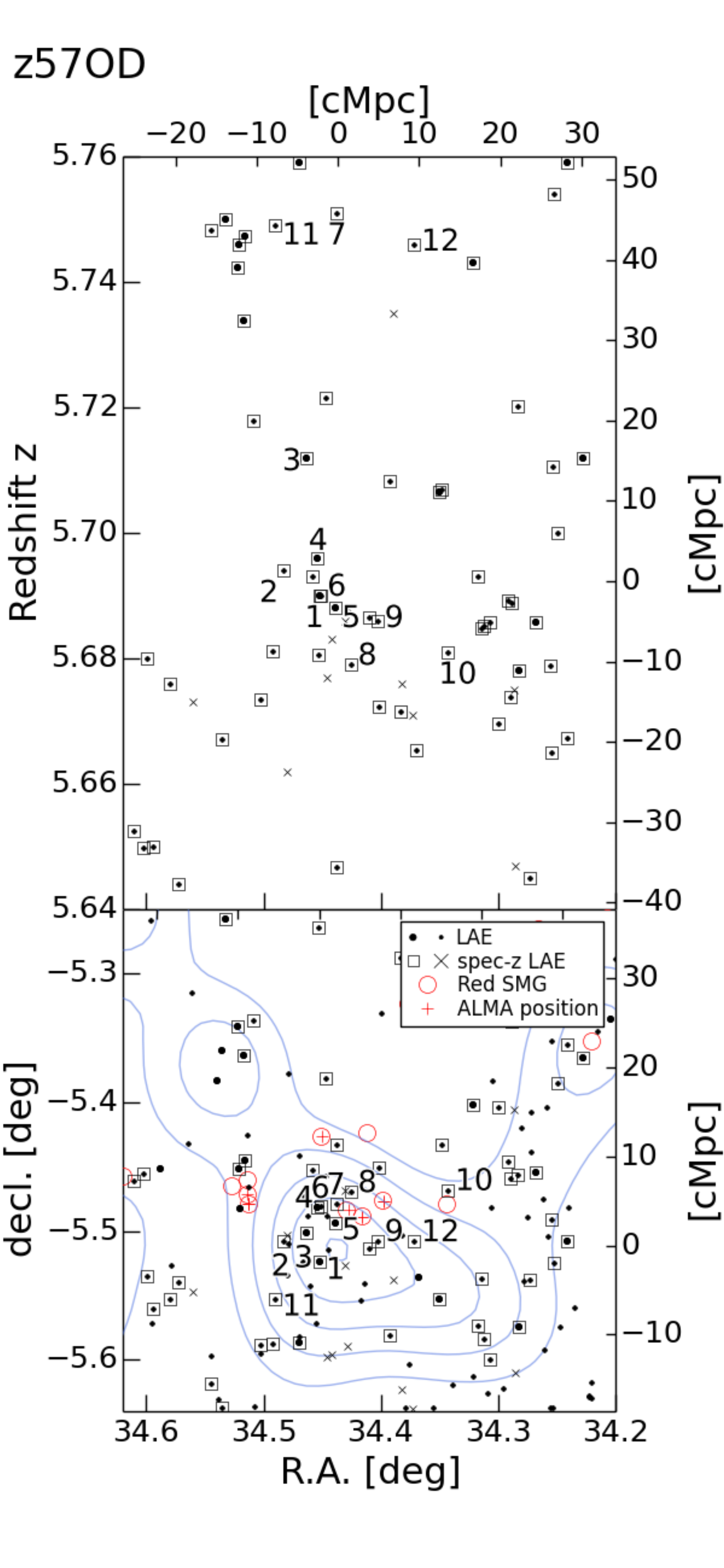}
 \end{center}
 \end{minipage}
 \begin{minipage}{0.58\hsize}
 \begin{center}
  \includegraphics[clip,bb=0 10 570 500,width=1\hsize]{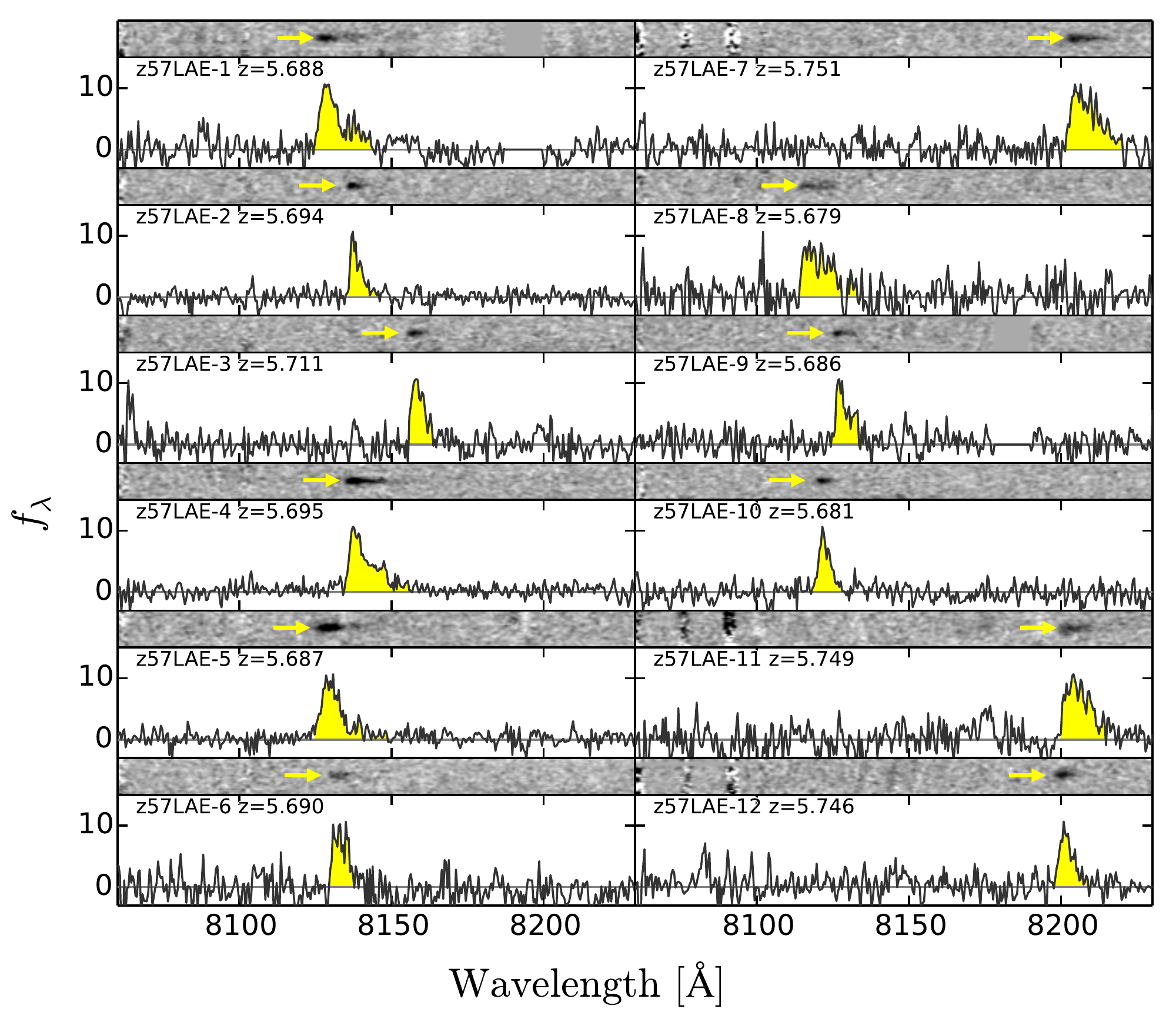}
 \end{center}
 \end{minipage}
 \end{center}
   \caption{
   {\bf Left panel:} 
   3D distribution of LAEs of z57OD.
   The large dots are LAEs whose NB magnitudes are brighter than 24.5.
   The LAEs indicated with the black squares are spectroscopically confirmed.
   The crosses are spectroscopically confirmed LAEs in \citet{2018NatAs.tmp..140J} but not identified in our photometric catalog.
   The numbers denote IDs of the LAEs. 
   The cyan contour shows the significance levels of the overdensity from $1\sigma$ to $5\sigma$.
   The red circles are the red SMGs (see Section \ref{ss_smg}), and the red crosses show the positions of the ALMA counterparts of the SMGs.
   {\bf Right panel:} 
   Examples of spectra of LAEs of z57OD. 
   The y-axes range of the 2D spectra are $\pm5\arcsec$.
   The y-axes in the 1D spectra are arbitrary.
    \label{fig_z57OD}}
\end{figure*}

\begin{figure*}
 \begin{center}
  \begin{minipage}{0.41\hsize}
 \begin{center}
  \includegraphics[clip,bb=0 40 430 915,width=1\hsize]{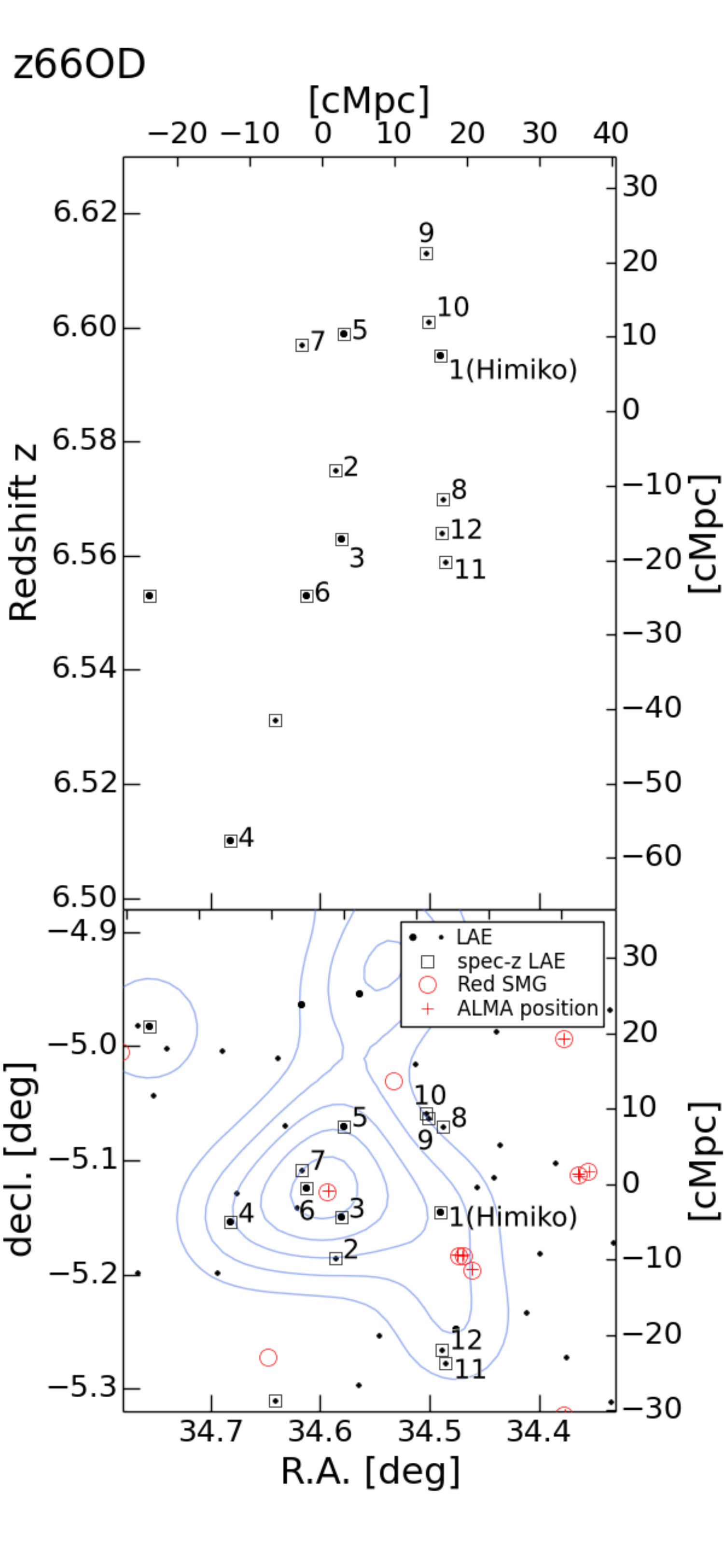}
 \end{center}
 \end{minipage}
 \begin{minipage}{0.58\hsize}
 \begin{center}
  \includegraphics[clip,bb=0 10 570 500,width=1\hsize]{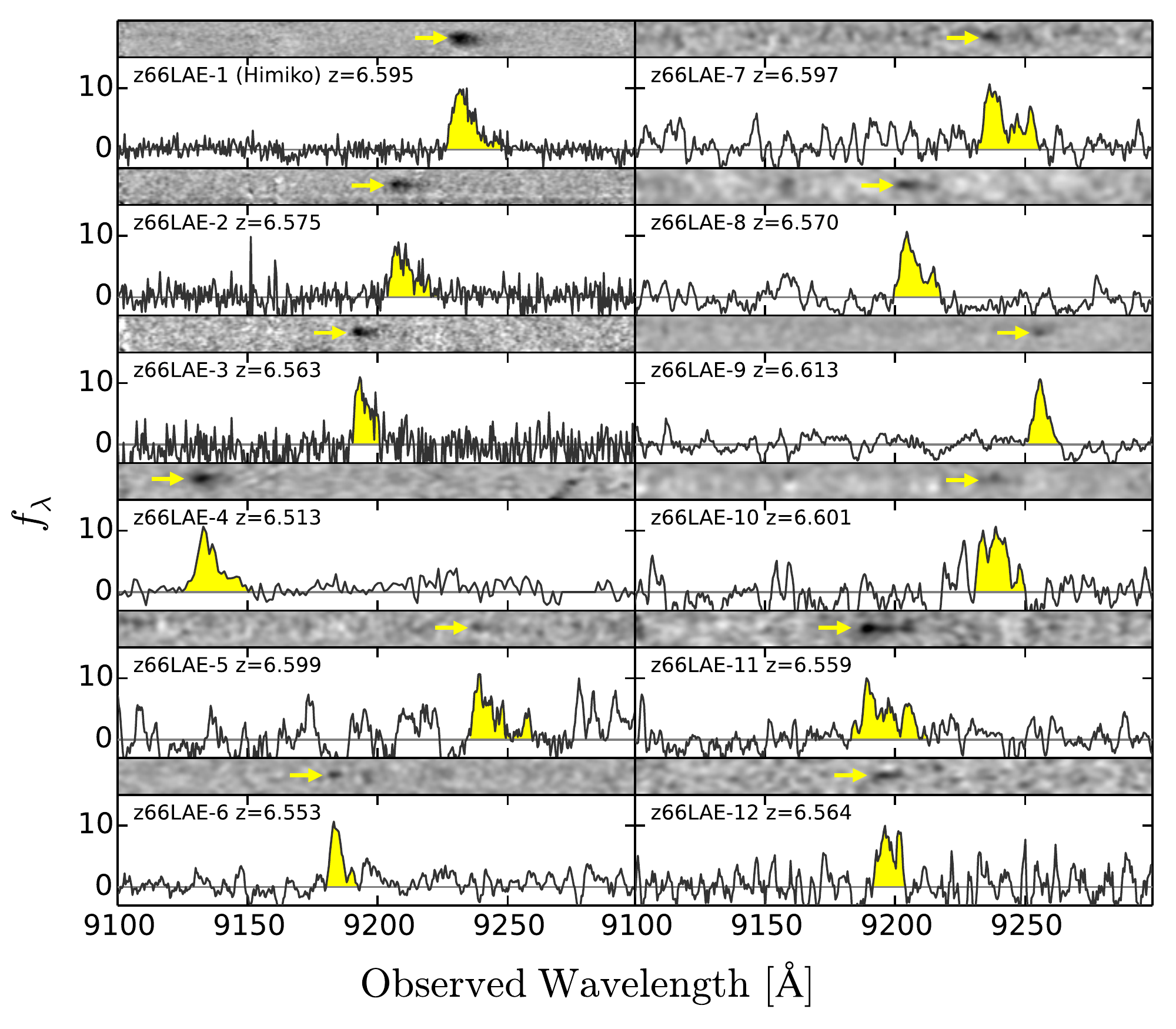}
 \end{center}
 \end{minipage}
 \end{center}
   \caption{
   Same as Figure \ref{fig_z57OD} but for z66OD.
   The large dots are LAEs whose NB magnitudes are brighter than 25.0.   
   The signals in the 2D spectra of z66LAE-4 ($\sim9270\ \m{\AA}$) and z66LAE-8 ($\sim9160\ \m{\AA}$) are residuals of the sky subtractions.
    \label{fig_z66OD}}
\end{figure*}

In the 3D maps, we identify z57OD ($z=5.692$) and z66OD ($z=6.585$) with 44 and 12 LAEs spectroscopically confirmed, respectively, which are located within $\sim1\sigma$ contours in Figures \ref{fig_z57OD} and \ref{fig_z66OD}.
\redc{The $1\sigma$ contours are roughly corresponding to the 20 cMpc-radius aperture.
According to theoretical studies in \citet{2017ApJ...844L..23C}, the 20 cMpc-radius aperture at $z\sim6$ includes $>90\%$ members of clusters at $z=0$.
We include z66LAE-8 located just outsize the $1\sigma$ contour, because it is within 20 cMpc from the center of z66OD.}
Figures \ref{fig_z57OD} and \ref{fig_z66OD} show the locations of LAEs, 2D projected contours, and spectra of the LAEs of z57OD and z66OD, respectively.
Tables \ref{tab_specz57} and \ref{tab_specz66} summarize properties of LAEs of z57OD and z66OD, respectively.
The average redshift of the LAEs of z66OD ($z=6.585$) suggests that z66OD is the most distant overdensity with $>10$ galaxies spectroscopically confirmed to date \citep[c.f., 3 galaxies at $z=7.1$ in][]{2018ApJ...863L...3C}.
\redc{Properties of overdensities in this work and in the literature are summarized in Table \ref{tab_PC}, which is based on objects listed in Table 5 in \citet{2013ApJ...779..127C} and new objects discovered since.}

\begin{deluxetable*}{cccccccc}
\tablecaption{Spectroscopically Confirmed LAEs of z57OD}
\tablehead{\colhead{ID} & \colhead{$\mathrm{R.A.\ (J2000)}$} & \colhead{$\mathrm{Decl.\ (J2000)}$} & \colhead{$z_\m{spec}$} & \colhead{$\m{log}L_\m{Ly\alpha}$} & \colhead{$M_\m{UV}$} & \colhead{$EW_\m{Ly\alpha}^0$} & \colhead{Ref.} \\
\colhead{(1)}& \colhead{(2)}& \colhead{(3)}& \colhead{(4)} &  \colhead{(5)}& \colhead{(6)}& \colhead{(7)}& \colhead{(8)}}
\startdata
z57LAE-1 & 02:17:48.46 & $-$05:31:27.02 & 5.688 & $43.06^{+0.04}_{-0.05}$ & $-20.9^{+0.3}_{-0.2}$ & $54^{+22}_{-13}$ & O08\\
z57LAE-2 & 02:17:55.83 & $-$05:30:26.94 & 5.694 & $42.57^{+0.10}_{-0.13}$ & $-18.9^{+1.1}_{-1.1}$ & $86^{+162}_{-38}$ & Hi18\\
z57LAE-3 & 02:17:51.14 & $-$05:30:03.64 & 5.711 & $42.74^{+0.08}_{-0.10}$ & $-19.6^{+0.8}_{-0.7}$ & $86^{+103}_{-43}$ & O08\\
z57LAE-4 & 02:17:49.11 & $-$05:28:54.17 & 5.695 & $43.17^{+0.03}_{-0.04}$ & $-19.8^{+0.8}_{-0.6}$ & $193^{+206}_{-83}$ & O08\\
z57LAE-5 & 02:17:45.24 & $-$05:29:36.01 & 5.687 & $43.09^{+0.04}_{-0.04}$ & $>-19.4$ & $>216$ & O08\\
z57LAE-6 & 02:17:48.19 & $-$05:28:51.92 & 5.690 & $42.59^{+0.09}_{-0.12}$ & $-18.9^{+1.1}_{-1.1}$ & $99^{+200}_{-49}$ & Hi18\\
z57LAE-7 & 02:17:45.01 & $-$05:28:42.37 & 5.751 & $42.71^{+0.09}_{-0.11}$ & $-20.7^{+0.2}_{-0.2}$ & $30^{+12}_{-8}$ & O08\\
z57LAE-8 & 02:17:42.17 & $-$05:28:10.55 & 5.679 & $42.91^{+0.06}_{-0.07}$ & $-20.8^{+0.2}_{-0.2}$ & $46^{+16}_{-11}$ & Hi18\\
z57LAE-9 & 02:17:36.68 & $-$05:30:27.57 & 5.686 & $42.53^{+0.11}_{-0.14}$ & $-18.9^{+1.1}_{-1.1}$ & $88^{+156}_{-46}$ & Hi18\\
z57LAE-10 & 02:17:22.28 & $-$05:28:05.30 & 5.681 & $42.76^{+0.06}_{-0.08}$ & $-18.9^{+1.1}_{-1.1}$ & $151^{+228}_{-69}$ & Hi18\\
z57LAE-11 & 02:17:57.66 & $-$05:33:09.16 & 5.749 & $42.75^{+0.10}_{-0.12}$ & $-20.0^{+0.6}_{-0.6}$ & $66^{+89}_{-29}$ & Hi18\\
z57LAE-12 & 02:17:29.18 & $-$05:30:28.50 & 5.746 & $42.48^{+0.13}_{-0.19}$ & $-19.1^{+1.0}_{-1.0}$ & $66^{+124}_{-36}$ & Hi18\\
z57LAE-13 & 02:16:54.60 & $-$05:21:55.53 & 5.712 & $43.10^{+0.04}_{-0.04}$ & $-20.1^{+0.7}_{-0.5}$ & $127^{+129}_{-48}$ & Hi18\\
z57LAE-14 & 02:17:04.30 & $-$05:27:14.30 & 5.686 & $43.15^{+0.04}_{-0.04}$ & $-20.3^{+0.6}_{-0.4}$ & $119^{+102}_{-40}$ & Hi18\\
z57LAE-15 & 02:17:07.85 & $-$05:34:26.51 & 5.678 & $43.24^{+0.03}_{-0.03}$ & $-20.6^{+0.4}_{-0.3}$ & $113^{+64}_{-31}$ & Hi18\\
z57LAE-16 & 02:17:24.02 & $-$05:33:09.62 & 5.707 & $43.32^{+0.02}_{-0.02}$ & $-21.3^{+0.2}_{-0.2}$ & $75^{+20}_{-14}$ & Hi18\\
z57LAE-17 & 02:18:03.87 & $-$05:26:43.45 & 5.747 & $42.90^{+0.06}_{-0.07}$ & $>-19.4$ & $>136$ & Hi18\\
z57LAE-18 & 02:18:04.17 & $-$05:21:47.25 & 5.734 & $42.87^{+0.06}_{-0.08}$ & $-21.4^{+0.2}_{-0.2}$ & $23^{+7}_{-5}$ & Hi18\\
z57LAE-19 & 02:18:05.17 & $-$05:27:04.06 & 5.746 & $42.89^{+0.06}_{-0.07}$ & $>-19.4$ & $>133$ & Hi18\\
z57LAE-20 & 02:18:05.28 & $-$05:20:26.89 & 5.742 & $42.80^{+0.08}_{-0.09}$ & $-20.5^{+0.5}_{-0.3}$ & $44^{+33}_{-16}$ & Hi18\\
z57LAE-21 & 02:18:28.87 & $-$05:14:23.01 & 5.737 & $43.38^{+0.02}_{-0.02}$ & $-20.4^{+0.6}_{-0.4}$ & $198^{+161}_{-64}$ & Hi18\\
z57LAE-22 & 02:18:30.53 & $-$05:14:57.80 & 5.688 & $43.27^{+0.03}_{-0.03}$ & $-20.4^{+0.6}_{-0.4}$ & $154^{+124}_{-50}$ & Hi18\\
z57LAE-23 & 02:17:13.81 & $-$05:35:58.23 & 5.686 & $42.86^{+0.09}_{-0.11}$ & $-21.0^{+0.3}_{-0.3}$ & $33^{+15}_{-10}$ & Hi18\\
z57LAE-24 & 02:18:00.70 & $-$05:35:18.92 & 5.673 & $43.04^{+0.05}_{-0.06}$ & $-21.6^{+0.1}_{-0.1}$ & $28^{+6}_{-5}$ & Hi18\\
z57LAE-25 & 02:17:58.09 & $-$05:35:15.35 & 5.681 & $42.55^{+0.11}_{-0.14}$ & $-19.0^{+1.0}_{-1.0}$ & $82^{+134}_{-41}$ & Hi18\\
z57LAE-26 & 02:17:14.93 & $-$05:35:02.77 & 5.685 & $42.50^{+0.12}_{-0.17}$ & $-20.6^{+0.2}_{-0.2}$ & $20^{+10}_{-7}$ & Hi18\\
z57LAE-27 & 02:17:34.16 & $-$05:34:52.56 & 5.708 & $42.63^{+0.09}_{-0.11}$ & $-18.9^{+1.1}_{-1.1}$ & $105^{+221}_{-48}$ & Hi18\\
z57LAE-28 & 02:17:16.10 & $-$05:34:24.23 & 5.693 & $42.73^{+0.08}_{-0.10}$ & $-19.1^{+1.1}_{-1.1}$ & $118^{+239}_{-59}$ & Hi18\\
z57LAE-29 & 02:17:05.63 & $-$05:32:17.66 & 5.645 & $42.89^{+0.08}_{-0.09}$ & $-20.7^{+0.3}_{-0.3}$ & $48^{+27}_{-14}$ & Hi18\\
z57LAE-30 & 02:17:15.53 & $-$05:32:14.04 & 5.685 & $42.51^{+0.11}_{-0.15}$ & $-19.9^{+0.4}_{-0.4}$ & $38^{+31}_{-15}$ & Hi18\\
z57LAE-31 & 02:17:38.28 & $-$05:30:48.70 & 5.687 & $42.86^{+0.07}_{-0.09}$ & $-19.9^{+0.6}_{-0.6}$ & $85^{+90}_{-33}$ & Hi18\\
z57LAE-32 & 02:17:01.13 & $-$05:29:28.40 & 5.665 & $42.53^{+0.12}_{-0.17}$ & $-19.1^{+1.1}_{-1.1}$ & $72^{+141}_{-39}$ & Hi18\\
z57LAE-33 & 02:17:09.50 & $-$05:27:31.49 & 5.674 & $42.71^{+0.08}_{-0.10}$ & $-19.1^{+1.1}_{-1.1}$ & $125^{+235}_{-66}$ & Hi18\\
z57LAE-34 & 02:17:07.96 & $-$05:27:23.16 & 5.720 & $42.52^{+0.12}_{-0.17}$ & $-19.1^{+1.1}_{-1.1}$ & $68^{+136}_{-36}$ & Hi18\\
z57LAE-35 & 02:17:49.99 & $-$05:27:08.07 & 5.693 & $43.08^{+0.06}_{-0.07}$ & $-20.3^{+0.6}_{-0.6}$ & $104^{+116}_{-40}$ & O08\\
z57LAE-36 & 02:17:36.38 & $-$05:27:01.62 & 5.672 & $43.16^{+0.04}_{-0.05}$ & $-20.2^{+0.5}_{-0.5}$ & $136^{+105}_{-47}$ & Hi18\\
z57LAE-37 & 02:17:09.95 & $-$05:26:46.53 & 5.689 & $42.91^{+0.07}_{-0.09}$ & $-19.4^{+1.1}_{-1.1}$ & $126^{+230}_{-58}$ & Hi18\\
z57LAE-38 & 02:17:45.19 & $-$05:25:57.75 & 5.647 & $42.59^{+0.11}_{-0.15}$ & $-19.7^{+0.6}_{-0.6}$ & $56^{+68}_{-26}$ & Hi18\\
z57LAE-39 & 02:16:59.94 & $-$05:23:05.33 & 5.700 & $42.49^{+0.12}_{-0.16}$ & $-19.8^{+0.5}_{-0.5}$ & $40^{+40}_{-17}$ & Hi18\\
z57LAE-40 & 02:16:57.88 & $-$05:21:16.99 & 5.667 & $43.16^{+0.04}_{-0.04}$ & $-19.7^{+0.8}_{-0.8}$ & $210^{+311}_{-89}$ & Hi18\\
z57LAE-41 & 02:18:02.18 & $-$05:20:11.48 & 5.718 & $42.59^{+0.09}_{-0.12}$ & $-18.9^{+1.1}_{-1.1}$ & $99^{+167}_{-49}$ & Hi18\\
z57LAE-42 & 02:17:01.43 & $-$05:18:41.68 & 5.679 & $42.71^{+0.08}_{-0.09}$ & $-19.0^{+1.1}_{-1.1}$ & $118^{+202}_{-55}$ & Hi18\\
z57LAE-43 & 02:17:00.61 & $-$05:31:30.27 & 5.754 & $42.56^{+0.11}_{-0.14}$ & $-20.0^{+0.5}_{-0.5}$ & $44^{+39}_{-18}$ & J18\\
z57LAE-44 & 02:17:52.63 & $-$05:35:11.79 & 5.759 & $43.49^{+0.02}_{-0.02}$ & $-22.1^{+0.1}_{-0.1}$ & $50^{+6}_{-5}$ & J18
\enddata
\tablecomments{(1) Object ID.
(2) Right ascension.
(3) Declination.
(4) Spectroscopic redshift of the Ly$\m{\alpha}$ emission line.
(5) Ly$\alpha$ luminosity in units of $\m{erg\ s^{-1}}$.
(6) Absolute UV magnitude or its $2\sigma$ lower limit in units of $\m{ABmag}$.
(7) Rest-frame Ly$\m{\alpha}$ EW or its $2\sigma$ lower limit in units of $\m{\AA}$.
(8) Reference (O08:\citealt{2008ApJS..176..301O}, Hi18:\citealt{2018arXiv180100531H}, J18:\citealt{2018NatAs.tmp..140J}).}
\label{tab_specz57}
\end{deluxetable*}

\begin{deluxetable*}{cccccccc}
\tablecaption{Spectroscopically Confirmed LAEs of z66OD}
\tablehead{\colhead{ID} & \colhead{$\mathrm{R.A.\ (J2000)}$} & \colhead{$\mathrm{Decl.\ (J2000)}$} & \colhead{$z_\m{spec}$} & \colhead{$\m{log}L_\m{Ly\alpha}$} & \colhead{$M_\m{UV}$} & \colhead{$EW_\m{Ly\alpha}^0$} & \colhead{Ref.} \\
\colhead{(1)}& \colhead{(2)}& \colhead{(3)}& \colhead{(4)} &  \colhead{(5)}& \colhead{(6)}& \colhead{(7)}& \colhead{(8)}}
\startdata
z66LAE-1 & 02:17:57.58 & $-$05:08:44.64 & 6.595 & $43.48^{+0.02}_{-0.03}$ & $-21.4^{+0.6}_{-0.4}$ & $91^{+68}_{-29}$ & O10\\
z66LAE-2 & 02:18:20.69 & $-$05:11:09.88 & 6.575 & $42.96^{+0.07}_{-0.09}$ & $>-19.9$ & $>59$ & O10\\
z66LAE-3 & 02:18:19.39 & $-$05:09:00.65 & 6.563 & $42.95^{+0.07}_{-0.09}$ & $-20.8^{+0.8}_{-0.6}$ & $49^{+60}_{-25}$ & O10\\
z66LAE-4 & 02:18:43.62 & $-$05:09:15.63 & 6.513 & $43.04^{+0.06}_{-0.08}$ & $-22.0^{+0.3}_{-0.2}$ & $20^{+10}_{-6}$ & Ha18\\
z66LAE-5 & 02:18:18.73 & $-$05:04:12.96 & 6.599 & $42.98^{+0.07}_{-0.08}$ & $-20.9^{+0.8}_{-0.6}$ & $50^{+59}_{-25}$ & This work\\
z66LAE-6 & 02:18:27.00 & $-$05:07:26.89 & 6.553 & $42.99^{+0.06}_{-0.08}$ & $>-20.5$ & $>66$ & This work\\
z66LAE-7 & 02:18:27.95 & $-$05:06:29.89 & 6.597 & $42.76^{+0.16}_{-0.26}$ & $-21.8^{+0.6}_{-0.6}$ & $14^{+27}_{-8}$ & This work\\
z66LAE-8 & 02:17:56.99 & $-$05:04:14.33 & 6.570 & $42.85^{+0.09}_{-0.12}$ & $>-20.1$ & $>59$ & This work\\
z66LAE-9 & 02:18:00.79 & $-$05:03:30.25 & 6.613 & $42.43^{+0.24}_{-0.57}$ & $>-20.2$ & $>19$ & This work\\
z66LAE-10 & 02:18:00.23 & $-$05:03:46.73 & 6.601 & $42.60^{+0.14}_{-0.21}$ & $>-20.0$ & $>42$ & This work\\
z66LAE-11 & 02:17:56.42 & $-$05:16:37.96 & 6.559 & $42.76^{+0.13}_{-0.19}$ & $-21.1^{+0.8}_{-0.8}$ & $22^{+74}_{-13}$ & This work\\
z66LAE-12 & 02:17:57.30 & $-$05:15:56.27 & 6.564 & $42.52^{+0.20}_{-0.39}$ & $>-20.2$ & $>32$ & This work
\enddata
\tablecomments{(1) Object ID.
(2) Right ascension.
(3) Declination.
(4) Spectroscopic redshift of the Ly$\m{\alpha}$ emission line.
(5) Ly$\alpha$ luminosity in units of $\m{erg\ s^{-1}}$.
(6) Absolute UV magnitude or its $2\sigma$ lower limit in units of $\m{ABmag}$.
(7) Rest-frame Ly$\m{\alpha}$ EW or its $2\sigma$ lower limit in units of $\m{\AA}$.
(8) Reference (O10:\citealt{2010ApJ...723..869O}, Ha18:\citealt{2018ApJ...859...84H}).}
\label{tab_specz66}
\end{deluxetable*}

Both z57OD and z66OD are located in the filamentary structures made by LAEs around these overdensities, extending over $40\ \m{cMpc}$.
We evaluate the extension of these overdensities in the redshift direction by calculating velocity dispersions of LAEs.
We select LAEs within $0.07\ \m{deg}$ from the centers (defined as the highest density peaks) of z57OD and z66OD, and calculate the rms of their velocities as velocity dispersions.
The calculated velocity dispersions are $1280\pm220\ \m{km\ s^{-1}}$ and $670\pm200\ \m{km\ s^{-1}}$, respectively, similar to the value of galaxies in overdensities found in \citet[][$1038\pm178\ \m{km\ s^{-1}}$]{2017arXiv170310170L} and \citet[][$647\pm124\ \m{km\ s^{-1}}$]{2012ApJ...750..137T}, respectively.
These velocity dispersions are compared with simulations in Section \ref{ss_sim}.

\citet{2018NatAs.tmp..140J} identify SXDS\_gPC in their spectroscopic survey.
Since the coordinate and redshift of SXDS\_gPC are the same as those of z57OD, we conclude that SXDS\_gPC is the same structure as z57OD.
\citet{2018NatAs.tmp..140J} spectroscopically confirm 46 LAEs at $z=5.7$ in the UD-SXDS field. 
34 LAEs among the 46 LAEs overlap with our LAE catalog, and traces similar large scale structures to the ones we identify.
However, the overdensity value and its significance ($\delta=5.6$, $\sim5\sigma$) are different from our measurements ($\delta=15.0$, $8.4\sigma$).
This is because the aperture size and magnitude limit of LAEs for the $\delta$ calculation are different between our measurements (10 cMpc-radius circular aperture and 24.5 mag) and Jiang et al. (35$^2$ cMpc$^2$ aperture and 25.5 mag).
\redc{If we calculate by adopting the same aperture size and magnitude limit as \citet{2018NatAs.tmp..140J} for spectroscopically confirmed LAEs, we obtain $\delta=4.8$ ($4.1\sigma$), comparable to the measurements of \citet{2018NatAs.tmp..140J}.}

\subsection{Comparison with Simulations}\label{ss_sim}
We compare our results with numerical simulations of \citet{2018PASJ...70...55I} to estimate halo masses of z57OD and z66OD. 
\citet{2018PASJ...70...55I} use $N$-body simulations with $4096^3$ dark matter particles in a comoving box of $162\ \m{Mpc}$.
The particle mass is $2.46\times10^6\ \m{M_\odot}$ and the minimum halo mass is $9.80\times10^7\ \m{M_\odot}$.
Halos' ionizing emissivity and IGM {\sc Hi} clumpiness are produced by a RHD simulation with a 20 comoving Mpc$^3$ box (Hasegawa et al. in prep.).
LAEs have been modeled with physically motivated analytic recipes as a function of halo mass.
LAEs are modeled based on the radiative transfer calculations by a radiative hydrodynamic simulation (Hasegawa et al in prep.).
In this work, we use the LAE model G with the late reionization history, which reproduces all observational results, namely the neutral hydrogen fraction measurements, Ly$\alpha$ luminosity functions, LAE angular correlation functions, and Ly$\alpha$ fractions in LBGs at $z\gtrsim6$.
Thus we expect that similar systems to z57OD and z66OD are found in the simulations.
We slice the $162\times162\times162\ \m{cMpc^3}$ box into four slices of $162\times162\times40.5\ \m{cMpc^3}$ whose depth ($\sim40\ \m{cMpc}$) is comparable to the redshift range of the narrowband selected LAEs at $z=5.7$ and $6.6$.
Magnitudes of the LAEs are calculated based on the transmission curves of the HSC filters.

\begin{figure*}
\begin{center}
  \includegraphics[clip,bb=0 0 680 380,width=1\hsize]{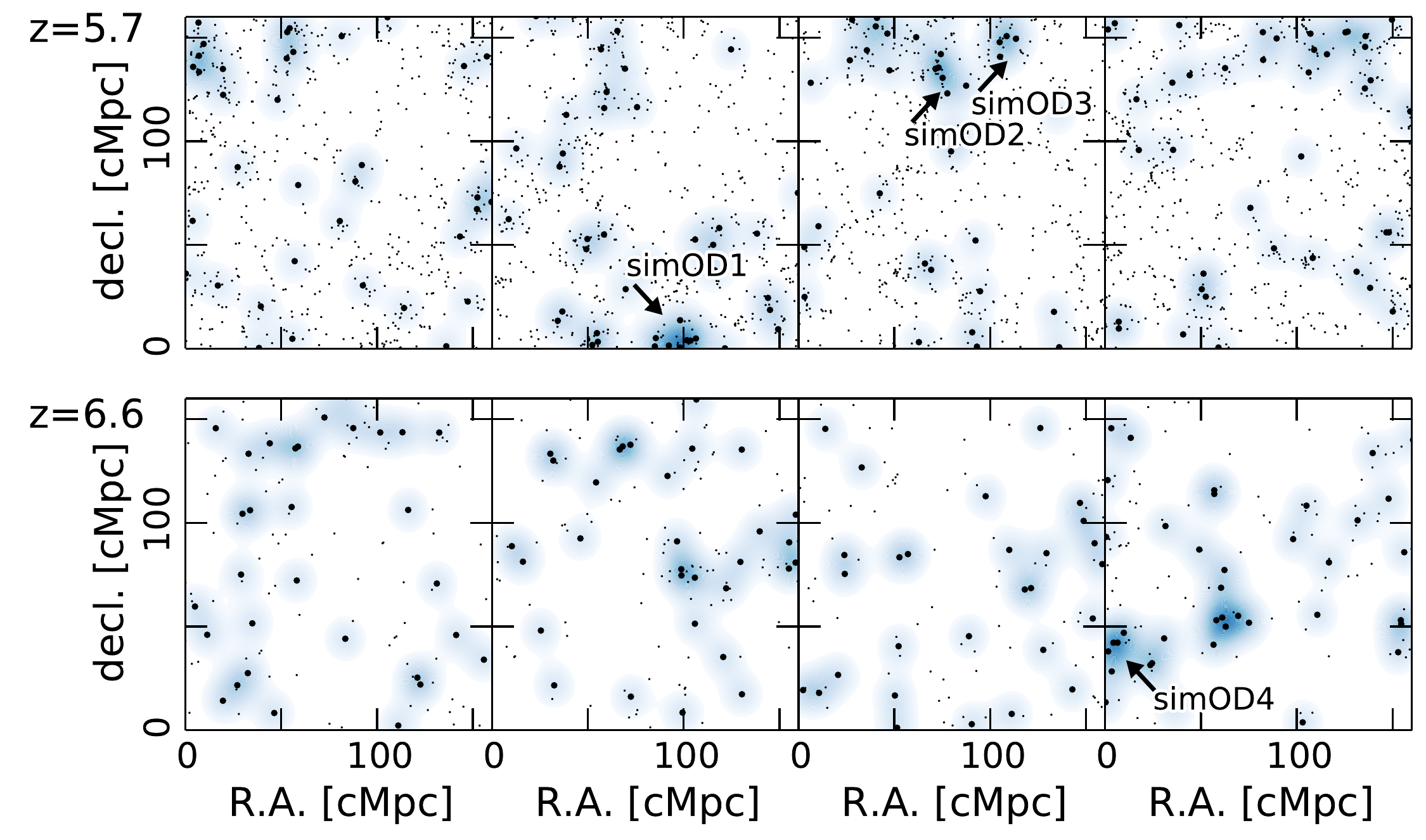}
 \end{center}
   \caption{
   {\bf Upper panels:} 2D map of LAEs at $z=5.7$ for four slices in the simulation box.
   Each slice has the 40 cMpc depth corresponding to the narrowband width.
   The black dots show the positions of the LAEs with $NB816<25.5\ \m{mag}$.
   The large dots are LAEs brighter than $NB816<24.5\ \m{mag}$.
   {\bf Lower panels:} Same as the upper panels but at $z=6.6$.
   The large dots are LAEs brighter than $NB921<25.0\ \m{mag}$.
   \label{fig_sim}}
\end{figure*}

We select $z=5.7$ and $6.6$ mock LAEs with $i-NB816>1.2$ and $z-NB921>1.0$, which are the same as our color criteria of Equations (\ref{eq_colorz57}) and (\ref{eq_colorz66}), respectively.
Then we use mock LAEs brighter than $NB816<24.5\ \m{mag}$ and $NB921<25.0\ \m{mag}$ at $z=5.7$ and $6.6$, respectively, and calculate the galaxy overdensity in each slice with a cylinder whose depth and radius are $40\ \m{cMpc}$ and $10\ \m{cMpc}$, respectively.
\redc{The average number densities of LAEs in the cylinder are $\overline{n}=0.39$ and $0.32$ at $z=5.7$ and $6.6$, respectively, which agree with observations within $1\sigma$ fluctuations.}
We show the calculated overdensity in each slice in Figure \ref{fig_sim}.
We define overdensities as regions whose overdensity significances are higher than $4\sigma$.
We calculate velocity dispersions of LAEs in the overdensities, using LAEs within $10\ \m{cMpc}$ from the centers of the overdensities, similar aperture size to the one used in the velocity dispersion calculations for z57OD and z66OD.

\begin{figure}
\begin{center}
  \includegraphics[clip,bb=0 0 500 300,width=1\hsize]{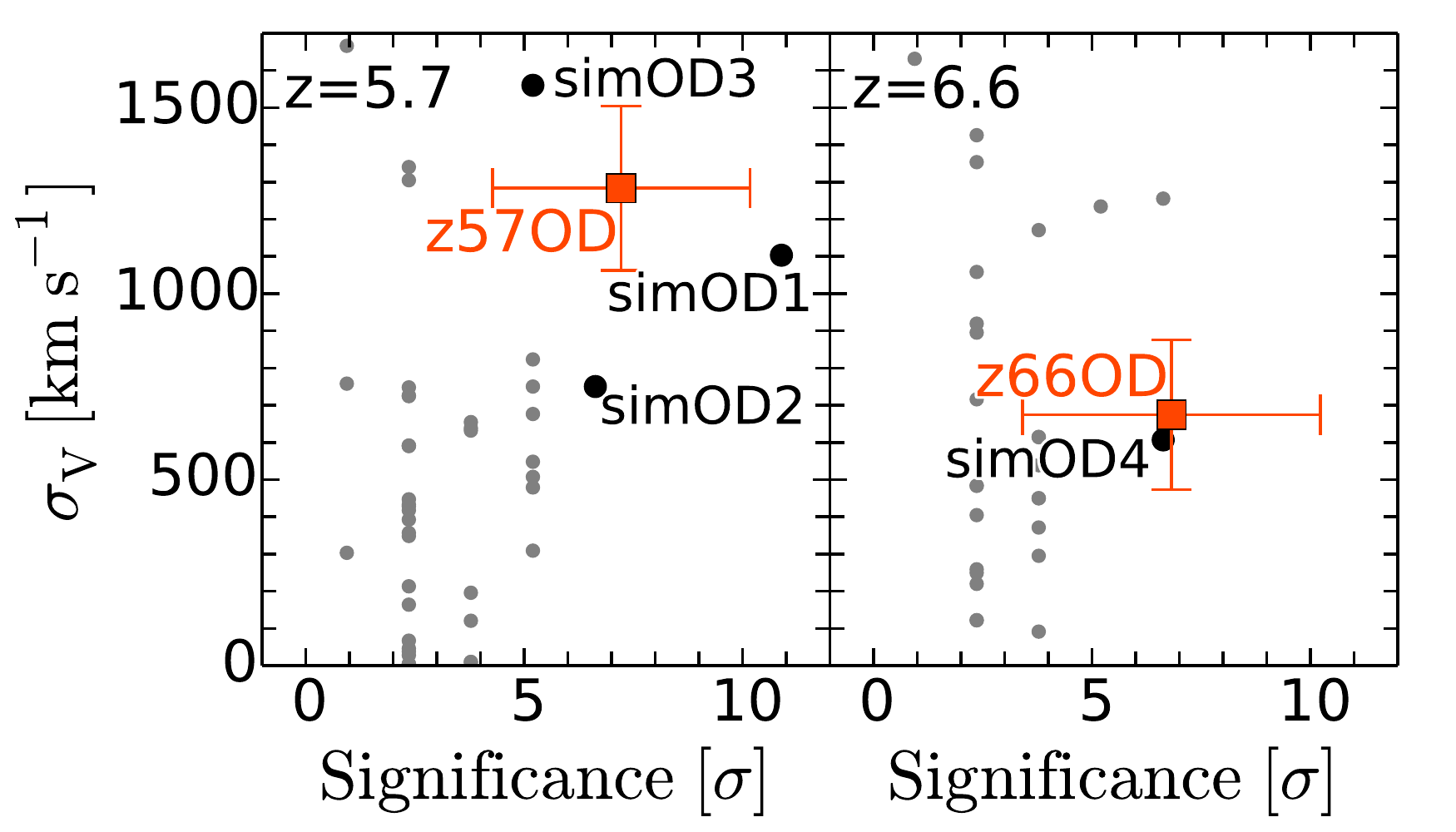}
 \end{center}
   \caption{
   Velocity dispersion of LAEs of overdensities as a function of the overdensity significance for LAEs at $z=5.7$ (left) and $z=6.6$ (right).
   The red squares show z57OD (left) and z66OD (right).
   The black and gray circles denote the overdensities identified in the simulations. 
   We identify three and one overdensities in simulations whose properties are similar to z57OD and z66OD, respectively.
   \label{fig_deltavsigma}}
\end{figure}

We compare the significances and velocity dispersions of the overdensities in the simulations with z57OD and z66OD in Figure \ref{fig_deltavsigma}.
At $z=5.7$, we find three overdensities, simOD1 \redc{($\delta=19.5$, $10.8\sigma$, $\sigma_\m{V}=1100\ \m{km\ s^{-1}}$)}, simOD2 \redc{($\delta=11.8$, $6.6\sigma$, $\sigma_\m{V}=750\ \m{km\ s^{-1}}$)}, and simOD3 \redc{($\delta=9.3$, $5.1\sigma$, $\sigma_\m{V}=1500\ \m{km\ s^{-1}}$)}, whose significance and velocity dispersion are comparable with z57OD with $\lesssim2\sigma$ uncertainties.
The masses of the most massive halos in these three overdensities are $1.0\times10^{12}\ \m{M_\odot}$, $4.7\times10^{11}\ \m{M_\odot}$, and $7.7\times10^{11}\ \m{M_\odot}$, respectively, at $z=5.7$.
At $z=6.6$, we identify one overdensity, simOD4 \redc{($\delta=13.7$, $7.3\sigma$, $\sigma_\m{V}=610\ \m{km\ s^{-1}}$)}, whose significance and velocity dispersion are comparable with z66OD with $\lesssim1\sigma$ uncertainties.
The mass of the most massive halo in simOD4 is $3.9\times10^{11}\ \m{M_\odot}$ at $z=6.6$.
Since the simulations do not go to $z\sim0$, we use the extended Press-Schechter model of \citet{2006MNRAS.369.1929H} to estimate the present-day halo masses of the $z=5.7$ and $6.6$ halos.
We find that these four overdensities in the simulations will grow to the cluster-scale halo ($M_\m{h}\sim 10^{14}\ \m{M_\odot}$) at $z\sim0$ with scatters of $\sim1\ \m{dex}$ in $M_\m{h}$, indicating that z57OD and z66OD are protoclusters.
Note that \cite{2009MNRAS.394..577O} reached the same conclusion on the progenitor of z57OD.

\redc{
We also estimate present-day halo masses of z57OD and z66OD using another method following previous studies \citep{1998ApJ...492..428S,2005A&A...431..793V,2012ApJ...750..137T}.
The halo mass at $z=0$ of a protocluster $M_\mathrm{h}$ is given by
\begin{equation}
M=\bar{\rho}V(1+\delta_\m{m}),
\end{equation}
where $\bar{\rho}=4.1\times10^{10}\ M_\odot\m{Mpc^{-3}}$ is the mean matter density of the universe, $V$ is the comoving volume of the protocluster that collapses into the cluster at $z=0$, and $\delta_\m{m}$ is the mass overdensity.
The mass overdensity $\delta_\m{m}$ is related to the galaxy overdensity $\delta$ with
\begin{equation}
1+b\delta_\m{m}=C(1+\delta),
\end{equation}
where $b$ is the bias factor of galaxies and $C$ is the correction factor for the redshift space distortion. 
We assume $C=1$ because this value is close to $1$ at high redshift \citep{1991MNRAS.251..128L}.
The biases of LAEs at $z=5.7$ and $6.6$ are estimated to be $b=4.1$ and $b=4.5$ in \citet{2018PASJ...70S..13O}.
Assuming $V=(4/3)\pi\times10^3\ \m{Mpc^3}$ \citep[typical size of a protocluster in ][]{2013ApJ...779..127C}, we estimate the present-day halo masses of z57OD and z66OD to be $4.8\times10^{14}\ M_\odot$ and $5.4\times10^{14}\ M_\odot$, which agree with simulations.
These estimated present-day halo masses support that z57OD and z66OD are protoclusters.
}

As discussed in the previous paragraph, we identify similar overdensities to z57OD in the simulation.
However, \citet{2018NatAs.tmp..140J} report that they do not find overdensities similar to z57OD in their cosmological simulation that is an update of a previous work \citep{2013ApJ...779..127C}.
This difference may be due to the different sizes of apertures used to search overdensities.
We use 10 cMpc-radius circular aperture, while \citet{2018NatAs.tmp..140J} use a larger, 35$^2$ cMpc$^2$ aperture.
Thus the simulations could reproduce overdensities in the small scale, while could not in the large scale.

\subsection{Correlation with Red SMGs}\label{ss_smg}
In Section \ref{ss_3d}, we identify the large scale structures made by LAEs, typically dust-poor star-forming galaxies.
It is important to investigate whether dust-obscured star-formation also traces the large scale structures.
We select high-redshift SMGs at $z\simeq 4-6$ (hereafter red SMGs) from the JCMT/SCUBA-2 Cosmology Legacy Survey $850\ \m{\mu m}$ source catalog \citep{2017MNRAS.465.1789G} \redc{using {\it Herschel}/SPIRE fluxes}.
It should be noted that $\sim850\ \m{\mu m}$ offers the negative K-correction to study SMGs with the same sensitivity at $z\sim6$ as at the $z=2-3$.

\redc{
To estimate {\it Herschel}/SPIRE fluxes and partially overcome confusion problem due to the large beam size, we apply de-blending approach by using higher resolution positional priors. 
We adopt positions of SCUBA-2 sources detected with $>4\sigma$ total noise and then apply simultaneous source-fitting routine available via SUSSEXtractor task in HIPE \citep{2010ASPC..434..139O,2007ApJ...661.1339S}.
The PSF of the JCMT/SCUBA-2 image is $14.\carcsec8$ \citep{2017MNRAS.465.1789G}.
The PSFs of the {\it Herschel}/SPIRE images are assumed to be Gaussian with FWHM being $17.\carcsec6$, $25.\carcsec1$ and $35.\carcsec2$ at $250\ \m{\mu m}$, $350\ \m{\mu m}$, and $500\ \m{\mu m}$ respectively.
Total flux uncertainties are estimated by quadratically adding the instrument and confusion noise. 
We further fully evaluate our selection via realistic end-to-end simulation based on galaxy model of \citet{2017A&A...607A..89B} which includes physical clustering based on abundance matching and galaxy-galaxy lensing.
Using this simulation, we simulate the exact criteria we applied on our real maps. 
The typical flux density error is 9 mJy at 500$\mu$m, which is in agreement with a value predicted by simulations.
}

\begin{figure*}
\begin{center}
 \begin{minipage}{0.47\hsize}
 \begin{center}
  \includegraphics[clip,bb=0 20 500 500,width=1\hsize]{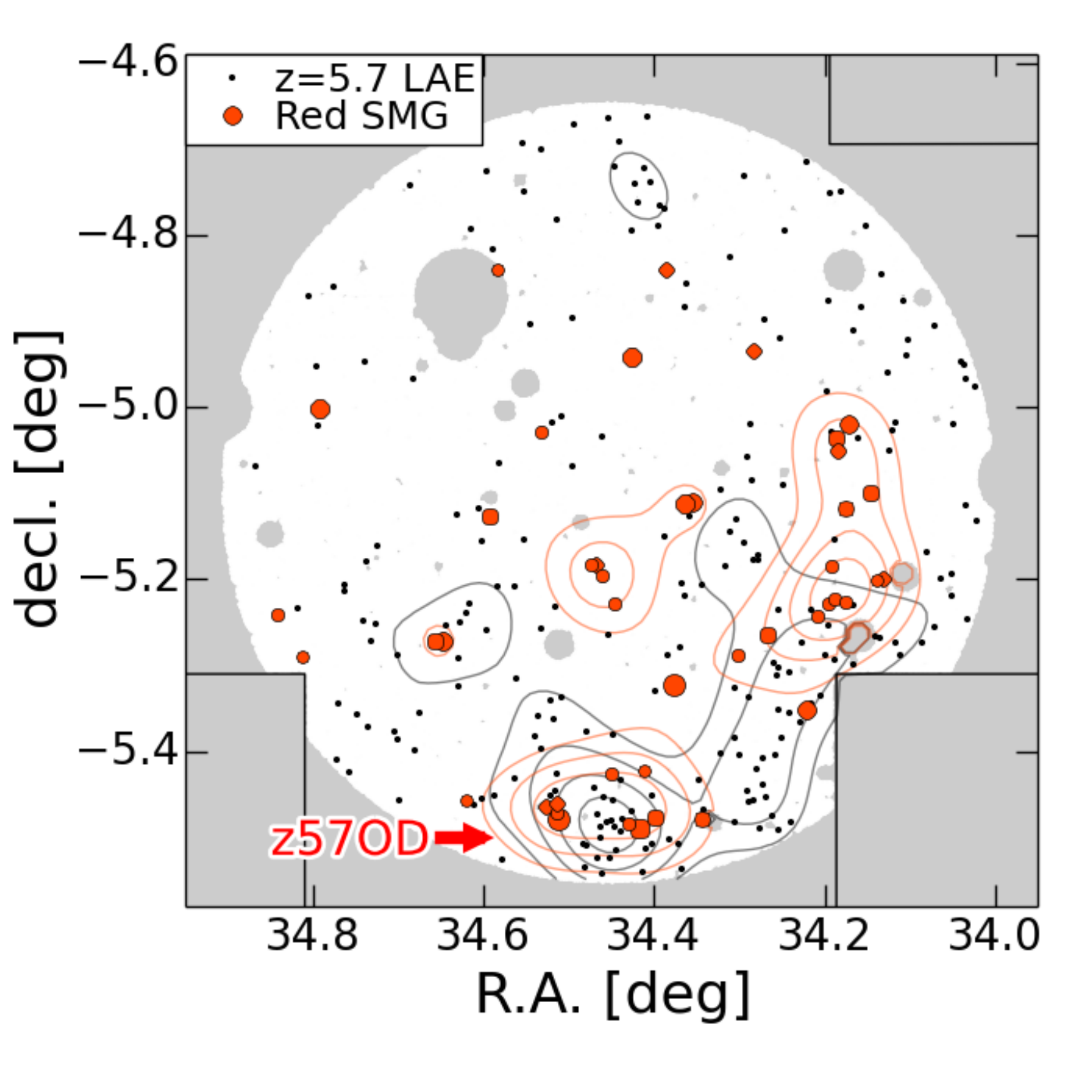}
 \end{center}
 \end{minipage}
  \begin{minipage}{0.51\hsize}
 \begin{center}
  \includegraphics[clip,bb=10 10 380 330,width=1\hsize]{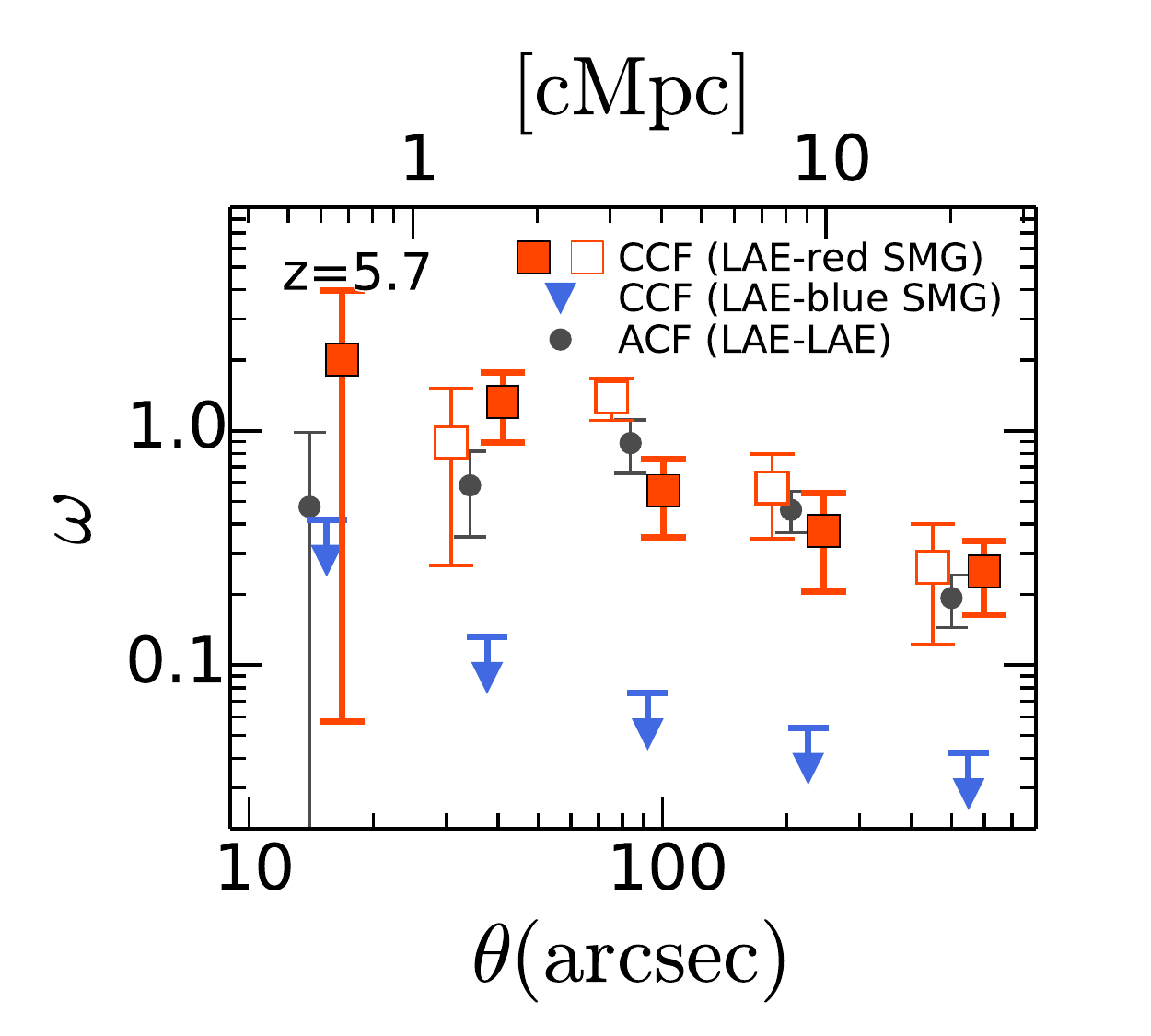}
 \end{center}
 \end{minipage}
 \end{center}
   \caption{
   {\bf Left panel:} Locations of the red-SMGs and LAEs at $z=5.7$.
   The red filled circles show the red SMGs and their sizes are scaled with the $850\ \m{\mu m}$ fluxes of the SMGs.
   The black circles are the LAEs at $z=5.7$, and the large circles show bright LAEs with $NB816<24.5$.
   \redc{The black and red contours shows the significance levels of the overdensity from $1\sigma$ to $4\sigma$ for $z=5.7$ LAEs and red SMGs, respectively.}
   {\bf Right panel:}
   Clustering of different popullations.
   The red filled (open) squares show the CCFs between the all (spectroscopically confirmed) LAEs at $z=5.7$ and red SMGs.
   The blue upper limits are the CCFs between the $z=5.7$ LAEs and the blue SMGs.
   The black circles show the ACFs of the $z=5.7$ LAEs for reference.
   We detect significant cross correlation signal between $z=5.7$ LAEs and red SMGs, indicating that a large number of the red SMGs are residing at $z=5.7$.
   \label{fig_ccf}}
\end{figure*}

To select red SMGs, we adopt the following criteria \citep{2018A&A...614A..33D}:
\begin{eqnarray}
S_\m{250\m{\mu m}}<S_\m{350\m{\mu m}}<S_\m{500\m{\mu m}} \label{eq_riser}
\end{eqnarray}
where $S_\m{250\m{\mu m}}$, $S_\m{350\m{\mu m}}$, and $S_\m{500\m{\mu m}}$ are the {\it Herschel} $250\ \m{\mu m}$, $350\ \m{\mu m}$, and $500\ \m{\mu m}$ fluxes, respectively.
\redc{Equation (\ref{eq_riser}) allows us to select $z\gtrsim4$ SMGs whose modified black body emission peak at $>500\ \m{\mu m}$ (see Figure 6 in \citealt{2018A&A...614A..33D}).\footnote{\redc{Although \citet{2018A&A...614A..33D} showed that most of the galaxies lie at $z<5$, this is because the number density of $z>5$ SMG is low \citep[e.g.,][]{2016ApJ...832...78I}.}}}
When using Equation (\ref{eq_riser}), we adopt the following three criteria to measure the {\it Herschel} colors correctly.
First, we use only sources whose $500\ \m{\mu m}$ fluxes are measured at $>2\sigma$ levels.
Second, if the sources are not detected in the $250\ \m{\mu m}$ and/or $350\ \m{\mu m}$ bands at the 2$\sigma$ levels, we replace fluxes with $2\sigma$ flux limits.
Third, we remove sources that are detected in $250\ \m{\mu m}$ but not in $350\ \m{\mu m}$.
After adopting these criteria and Equation (\ref{eq_riser}), we reduce low-redshift interlopers using ALMA and Subaru/HSC data.
We cross-match the SCUBA-2 sources with ALMA sources in archival data \citep[see also][]{2018ApJ...860..161S} within $10\arcsec$, and identify ALMA counterparts of the SCUBA-2 sources if present.
The ALMA data we use are taken in band 7, with typical $1\sigma$ noise level and angular resolution are $0.2\ \m{mJy/beam}$ and $0.\carcsec2$, respectively.
\redc{We identify ALMA counterparts of more than 70\% of the SCUBA-2 sources, and most of the rest are not observed with ALMA.}
We then measure fluxes at the positions of the ALMA counterparts in the HSC $g$ and $r$ images, and exclude SCUBA-2 sources with detection at $>3\sigma$ levels in the HSC $g$ or $r$ band images (bluer than the Lyman break at $z\simeq 4-6$).
Finally, we apply masks of diffraction spikes and halos from bright objects in the same fashion as for our LAEs, and obtain the final red SMG sample.
We also define SMGs not selected with above criteria as blue SMGs, which will be used for a null test.
In addition, we select LAEs at $z=5.7$ located in the sky coverage of the SCUBA-2 observation.
Finally we obtain 44 red SMGs, 673 blue SMGs, and 227 LAEs (77 spectroscopically confirmed).
Note that there is no overlap between the LAEs and the ALMA sources within $2\arcsec$.
Since LAEs are typically dust-poor weak 850 $\mu$m and [{\sc Cii}]158$\mu$m emitters \citep{2018ApJ...859...84H}, finding no overlap is reasonable.
\redc{According to \citet{2017MNRAS.465.1789G}, the false detection rate is $<6\%$ at the $>4\sigma$ detection.
Since we will test whether the red SMGs are at $z=5.7$ or not by the cross-correlation analysis later, we do not take this false detection rate into account here.}

\redc{The left panel in Figure \ref{fig_ccf} shows locations of the red SMGs and $z=5.7$ LAEs.
We find that some of the red SMGs are clustering around z57OD ($\m{R.A.}=34.26, \m{decl.}=-5.54$).}
We calculate the cross-correlation function (CCF) of the 227 LAEs at $z=5.7$ and the 44 red SMGs using the estimator in \citet{1993ApJ...412...64L}:
\begin{equation}
\omega(\theta)=\frac{D_1 D_2(\theta)-D_1 R_2(\theta)-R_1 D_2(\theta)+R_1 R_2 (\theta)}{R_1 R_2 (\theta)},
\end{equation}
where $DD$, $DR$, $RD$, and $RR$ are the numbers of galaxy-galaxy, galaxy-random, random-galaxy, and random-random pairs for the group 1 and 2.
We also calculate the CCF between the 77 spectroscopically confirmed LAEs and red SMGs, the CCF between the 227 LAEs and 775 blue SMGs, and angular auto-correlation functions (ACFs) of the 227 LAEs for reference.
Using SCUBA-2 SMGs may have the blending bias effect on the correlation function measurements due to confusion introduced by the coarse angular resolution \citep{2013MNRAS.432....2K,2018ApJ...860..161S}.
However, the effect is expected to be small, a factor of $\sim1.2-1.3$ \citep{2017MNRAS.469.3396C}.
We estimate statistical errors of the CCFs and ACF using the Jackknife estimator.
We divide the samples into 47 Jackknife subsamples of about $500^2\ \m{arcsec}^2$, comparable to the maximum angular size of the correlation function measurements.
Removing one Jackknife subsample at a time for each realization, we compute the covariance matrix as
\begin{equation}
C_{ij}=\frac{N-1}{N}\sum^{N}_{l=1}\left[\omega^l(\theta_i)-\bar{\omega}(\theta_i)\right]\left[\omega^l(\theta_j)-\bar{\omega}(\theta_j)\right].
\end{equation}
where $N$ is the total number of the Jackknife samples, and $\omega^l$ is the estimated CCFs or ACF from the $l$th realization.
$\bar{\omega}$ denotes the mean CCFs and ACF.
We apply a correction factor (typically $\sim1.1$) given by \citet{2007A&A...464..399H} to an inverse covariance matrix in order to compensate for the bias introduced by the statistical noise.

The calculated CCFs and ACF are presented in the right panel of Figure \ref{fig_ccf}.
We detect the signal of the cross-correlation between the LAEs at $z=5.7$ and red SMGs.
We evaluate the significance of the correlation by calculating the $\chi^2$ value,
\begin{equation}
\chi^2=\sum_{i.j}\left[\omega(\theta_i)-\omega_\m{model}(\theta_i)\right]C^{-1}_{i,j}\left[\omega(\theta_j)-\omega_\m{model}(\theta_j)\right],
\end{equation}
where $\omega_\m{model}=0$ for the non-detection case. 
We obtain $\chi^2=13.0$, indicating the $99.97\%$ significance correlation.
If we use the spectroscopically confirmed LAEs, the significance level of the cross-correlation is still $96\%$.
We do not detect the $>2\sigma$ correlation signal between the LAEs and blue SMGs\redc{, nor the LAEs and all SMGs}.
These significant correlations between the LAEs and red SMGs indicate that the red SMGs also trace the large scale structure with z57OD made by the LAEs, similar to the SSA-22 protocluster at $z=3.1$ \citep{2009Natur.459...61T,2014MNRAS.440.3462U}.
We also calculate cross correlation functions between the LAEs at $z=6.6$ and red SMGs, but do not detect a significant correlation signal beyond $2\sigma$.

We evaluate the fraction of the red SMGs located at $z=5.7$.
If all of the SMGs and LAEs are at $z=5.7$, the large scale ($\gtrsim1\ \m{cMpc}$) amplitude of the CCF between the LAEs and red SMGs is expressed as $b_\m{LAE}b_\m{SMG}\xi_\m{DM}$, where $b_\m{LAE}$, $b_\m{SMG}$, and $\xi_\m{DM}$ are the large scale bias of the LAE, the large scale bias of the SMG, and the dark matter correlation function.
If some of the red SMGs are not at $z=5.7$, the CCF amplitude will decrease by a factor of $1-f_\m{c}$, where $f_\m{c}$ is a fraction of the red SMGs that are not at $z=5.7$.
The large scale amplitude of the ACF of LAEs is $b_\m{LAE}^2\xi_\m{DM}$.
Because observed amplitude of the CCF between the LAEs and red SMGs are comparable to that of the ACF of LAEs, we get
\begin{equation}
b_\m{LAE}b_\m{SMG}\xi_\m{DM}(1-f_\m{c})=b_\m{LAE}^2\xi_\m{DM},
\end{equation}
and
\begin{equation}
(1-f_\m{c})=\frac{b_\m{LAE}}{b_\m{SMG}}.
\end{equation}
The large scale bias of LAEs at $z=5.7$ is typically $b_\m{LAE}\simeq4$ \citep{2018PASJ...70S..13O}.
The bias of SMGs is expected to be larger than that of LAEs ($b_\m{SMG}>b_\m{LAE}$), because SMGs are thought to be more massive than LAEs. 
For example, the large scale bias of SMGs is typically $\sim3$ times larger than that of LAEs at $z\sim2-3$ \citep[e.g.,][]{2003ApJ...582....6W,2007ApJ...671..278G,2009ApJ...707.1201W,2010ApJ...723..869O}.
On the other hand, the effective volume of our narrow-band data is $\sim200\times200\times80\ \m{cMpc^3}$.
Only one halo as massive as $M_\m{h}\sim10^{13}\ \m{M_\odot}$ is expected to exist in this volume, on average \citep{2008ApJ...688..709T}.
Thus we get the upper limit of the bias of the SMGs as $b_\m{SMG}<b(M_\m{h}=10^{13}\ \m{M_\odot})\simeq14$.
From the lower and upper limits obtained, $4<b_\m{SMG}<14$, we expect that the fraction of the red SMGs at $z=5.7$ are $\sim30-100\ \%$, suggesting that $\sim10-40$ red SMGs are at $z=5.7$.
This is higher than the expectation from the redshift distribution in \citet[][their Figure 7]{2018A&A...614A..33D}, hinting that large number of the red SMGs are clustering at $z=5.7$.
ALMA follow-up observations for these red SMGs are now being prepared.
It is interesting that the CCF shows strong correlation between the LAEs and the red SMGs even at the $<20\arcsec$ scale, while the ACF does not.
It indicates that LAE-red SMG pairs can be more easily found in the $<20\arcsec$ scale than LAE-LAE pairs.

\subsection{Star Formation Activity in z57OD and z66OD}

To understand star formation activities in z57OD and z66OD, we investigate spectral energy distributions (SEDs) of the LAEs of z57OD and z66OD.
We use the images of Subaru/HSC $grizyNB816NB921$, UKIRT/WFCAM $JHK$ in the UKIDSS/UDS project \citep{2007MNRAS.379.1599L}, and {\it Spitzer}/IRAC $[3.6]$ and $[4.5]$ bands in the SPLASH project (P. Capak in prep.).
Some LAEs are detected in the NIR images, and we can constrain SEDs of them.
Regarding LAEs not detected in the NIR images, we stack images of these LAEs, and make subsamples ("non detection stack" subsamples) in z57OD and z66OD.
We also stack images of all LAEs in z57OD and z66OD ("all stack" subsamples) to investigate averaged properties.

\begin{figure*}
\begin{center}
 \begin{minipage}{0.6\hsize}
 \begin{center}
  \includegraphics[clip,bb=10 0 570 300,width=1\hsize]{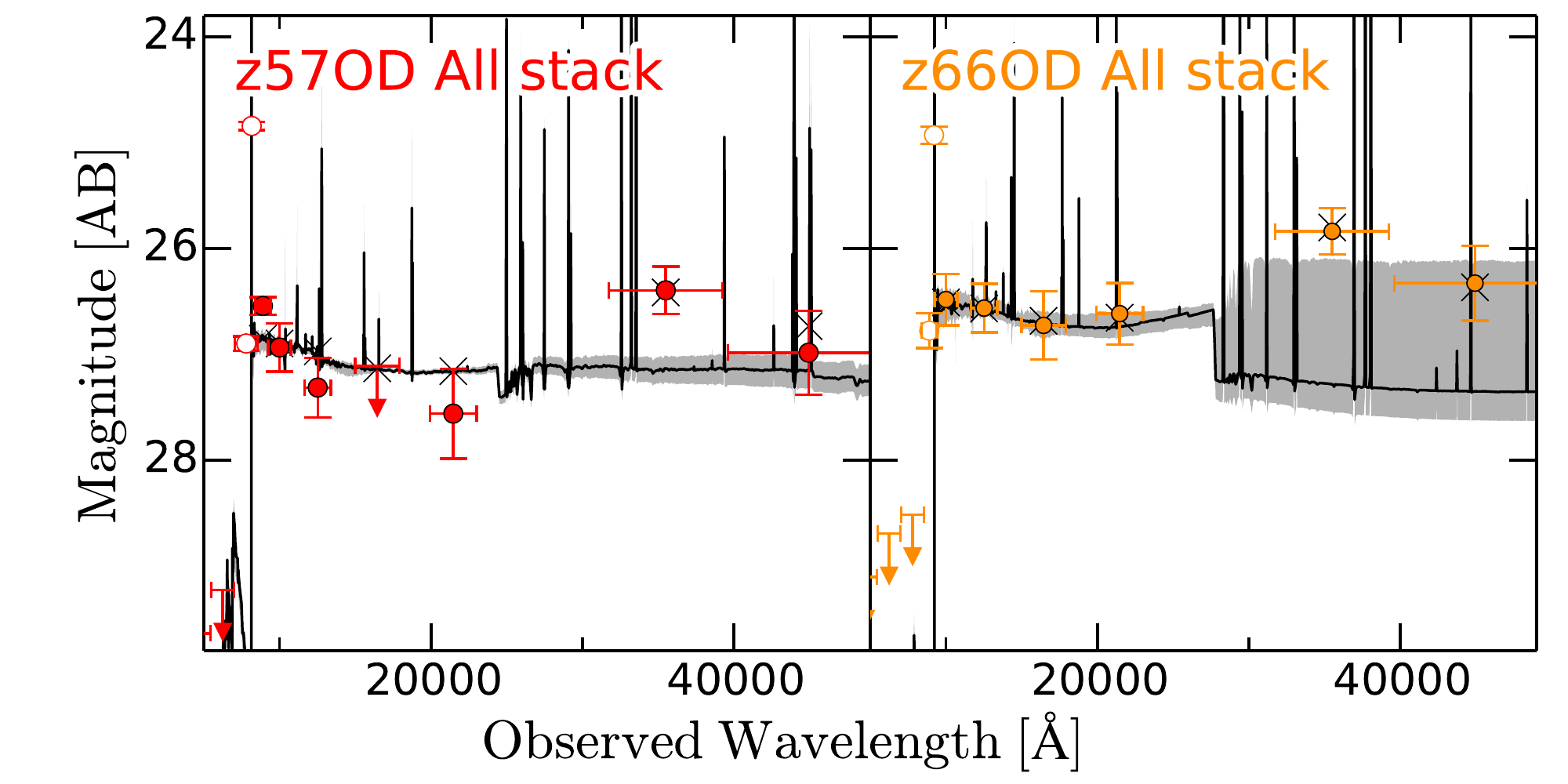}
 \end{center}
 \end{minipage}
  \begin{minipage}{0.39\hsize}
 \begin{center}
  \includegraphics[clip,bb=0 0 300 300,width=1\hsize]{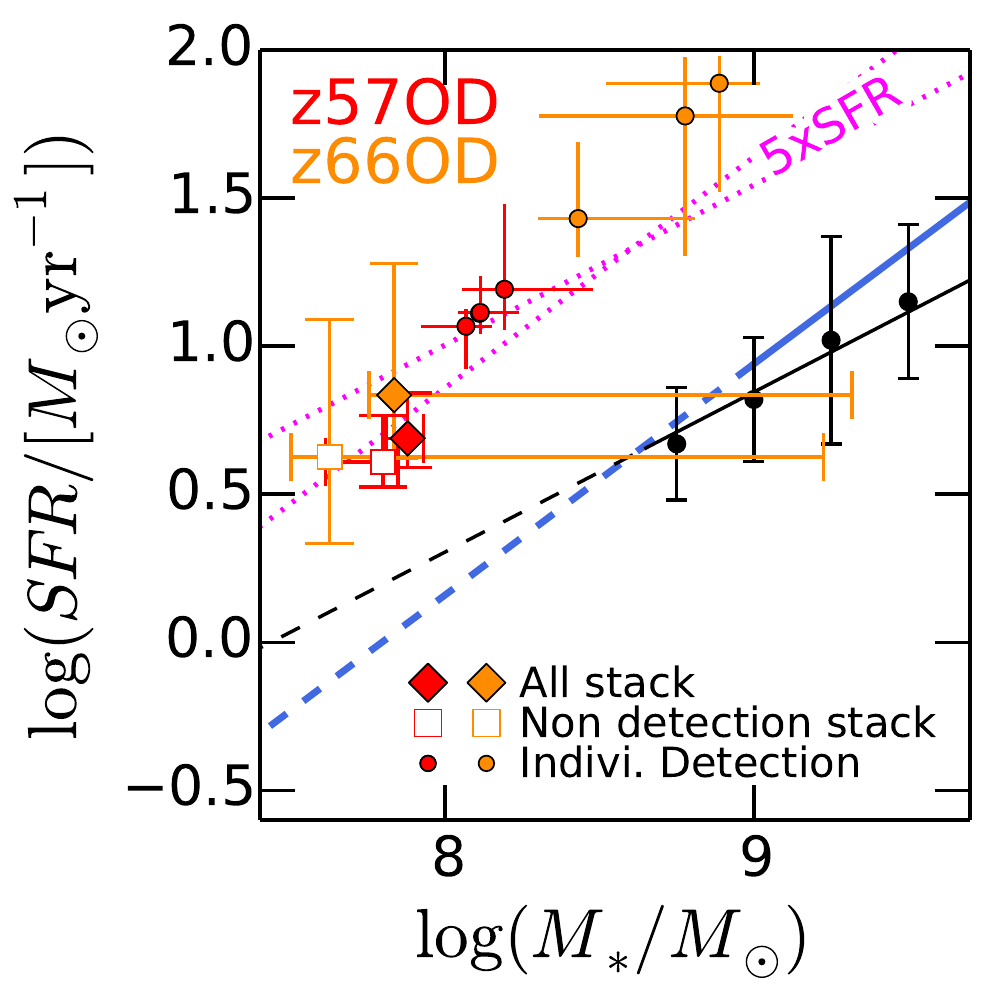}
 \end{center}
 \end{minipage}
 \end{center}
   \caption{
   {\bf Left and center panel:} 
   SEDs of the "all stack" subsamples in z57OD and z66OD.
      The circles represent the magnitudes in the stacked images of each subsample.
   The filled circles are magnitudes used in the SED fittings.
   We do not use the magnitudes indicated with the open circles which are affected by the IGM absorption.
   The dark gray lines with the gray circles show the best-fit model SEDs, and the light gray regions show the $1\sigma$ uncertainties of the best-fit model SEDs.
   {\bf Right panel:}
   SFRs of the LAEs in z57OD and z66OD as functions of the stellar mass.
   The red and orange diamonds (open squares) are SFRs of the all (non detection) stack subsamples of z57OD and z66OD, respectively.
   SFRs of the individual LAEs detected in the NIR images are shown with the red and orange circles for z57OD and z66OD, respectively.
   The black line with the circles and the blue lines show results of the star formation main sequence of \citet{2015ApJ...799..183S} at $z\sim6$ and \citet{2014ApJ...791L..25S} at $z=4.8-6.0$, respectively.
      The dashed lines represent extrapolations from the ranges these studies investigate.
      The SFRs of the LAEs in z57OD and z66OD are $\sim5$ times higher than galaxies in the main sequence.
   \label{fig_sed}}
\end{figure*}

Firstly we run {\tt T-PHOT} \citep{2016A&A...595A..97M} and generate residual IRAC images where only the LAEs under analysis are left.
As high-resolution prior images in the {\tt T-PHOT} run, we use HSC $grizyNB$ stacked images whose PSFs are $\sim0.\carcsec7$.
Then we visually inspect all of our LAEs and exclude sources due to the presence of bad residual features close to the targets that can possibly affect the photometry.
We cut out $12\arcsec\times12\arcsec$ images of the LAEs in each band, and generate median-stacked images of the subsamples in each bands with {\tt IRAF} task {\tt imcombine}.
We show the SEDs of the "all stack" subsamples at $z=5.7$ and $6.6$ in the left and center panels in Figure \ref{fig_sed}, respectively.

We generate the model SEDs at $z=5.7$ and $6.6$ using BEAGLE \citep{2016MNRAS.462.1415C}.
In BEAGLE, we use the combined stellar population and photoionization models presented in \citet{2016MNRAS.462.1757G}. 
Stellar emission is based on an updated version of the population synthesis code of \citet{2003MNRAS.344.1000B}, while gas emission is computed with the standard photoionization code CLOUDY \citep{2013RMxAA..49..137F} following the prescription of \citet{2001MNRAS.323..887C}.
The IGM absorption is considered following a model of \cite{2014MNRAS.442.1805I}.  
In BEAGLE we vary the total mass of stars formed, ISM metallicity ($Z_\m{neb}$), ionization parameter ($U_\m{ion}$), star formation history, stellar age, and $V$-band attenuation optical depth ($\tau_V$), while we fix the dust-to-metal ratio ($\xi_d$) to $0.3$ \citep[e.g.,][]{2017MNRAS.471.1743D}, and adopt the \citet{2000ApJ...533..682C} dust extinction curve.
We choose the constant star formation history because it reproduces SEDs of high redshift LAEs \citep{2010ApJ...724.1524O,2018ApJ...859...84H}.
The choice of the extinction law does not affect our conclusions, because our SED fittings infer dust-poor populations such as $\tau_V=0.0-0.1$.
We vary the four adjustable parameters of the model in vast ranges, $-2.0<\m{log}(Z_\m{neb}/Z_\odot)<0.2$ (with a step of $0.1\ \m{dex}$), $-3.0<\m{log}U_\m{ion}<-1.0$ (with a step of $0.1\ \m{dex}$), $6.0<\m{log}(\m{Age/yr})<9.0$ (with a step of $0.1\ \m{dex}$), and $\tau_\m{V}=[0,0.05,0.1,0.2,0.4,0.8,1.6,2]$.
We assume that the stellar metallicity is the same as the ISM metallicity, with interpolation of original templates. 
We fit our observed SEDs with these model SEDs, and derive stellar masses and SFRs of the subsamples and individuals.
In the "all stack" subsample at $z=5.7$, we can constrain the stellar mass, SFR, and metallicity.
In the other subsamples, we fix the metallicity to $\m{log}(Z/Z_\odot)=-0.6$ that is the best-fit value of the "all stack" subsample at $z=5.7$, because we cannot constrain the metallicity due to the poor signal-to-noise ratio.

In the right panel in Figure \ref{fig_sed}, we plot the measured stellar masses and SFRs for the LAEs of  z57OD and z66OD.
We compare them with the star formation main sequence that is determined with field LBGs.
All the subsamples including "all stack", "non detection stack", and individual galaxies show SFRs more than $\sim5$ times higher than the main sequence galaxies in the same stellar masses, indicating that the LAEs in z57OD and z66OD are actively forming stars.

We then calculate the SFR densities of z57OD and z66OD, and compare them with the cosmic average (a.k.a the Madau-Lilly plot).
\redc{We measure the SFR densities using observed galaxies located within $1$ physical Mpc (pMpc) from the centers of the overdensities, following previous studies \citep[e.g.,][]{2014MNRAS.439.1193C,2016MNRAS.460.3861K}.
We find that 16 LAEs and 3 red SMGs (5 LAEs and 1 red SMG) are within the 1 pMpc-radius aperture around z57OD (z66OD).}
For z57OD, we measure the total SFR density of the observed LAEs and red SMGS, because the cross correlation signal suggests that $30-100\%$ of the red SMGs trace the LAE large scale structures.
We assume that the average SFR of one LAE is $\sim10\ M_\odot\ \m{yr^{-1}}$ based on the SED fitting results. 
We calculate SFRs of the red SMGs from the $850\ \m{\mu m}$ fluxes assuming the redshift of $z=5.7$, the dust temperature of $T_\m{dust}=40\ \m{K}$ \citep{2013A&A...557A..95R,2017ApJ...847...21F}, and the emissivity index of $\beta=1.5$ \citep{2005ApJ...622..772C}.
\redc{The effect of these assumptions is not significant on our conclusions.
For example, the $\Delta T_\m{dust}=10\ \m{K}$ or $\Delta\beta=1.5$ difference changes the SFR density only by a factor of $<2$.
With this assumed temperature, the CMB effect is negligible \citep[$<5\%$;][]{2013ApJ...766...13D}.}
The uncertainty of the SFR density corresponds to the uncertainty of the fraction of the red SMGs residing at $z=5.7$ ($30-100\%$), because the total SFR is dominated by the SFRs of the red SMGs.

For z66OD, since we do not know whether the SMG are also at $z=6.6$, we calculate the lower limit of the SFR density considering only the LAEs. 

\begin{figure}
\begin{center}
  \includegraphics[clip,bb=0 0 300 300,width=0.8\hsize]{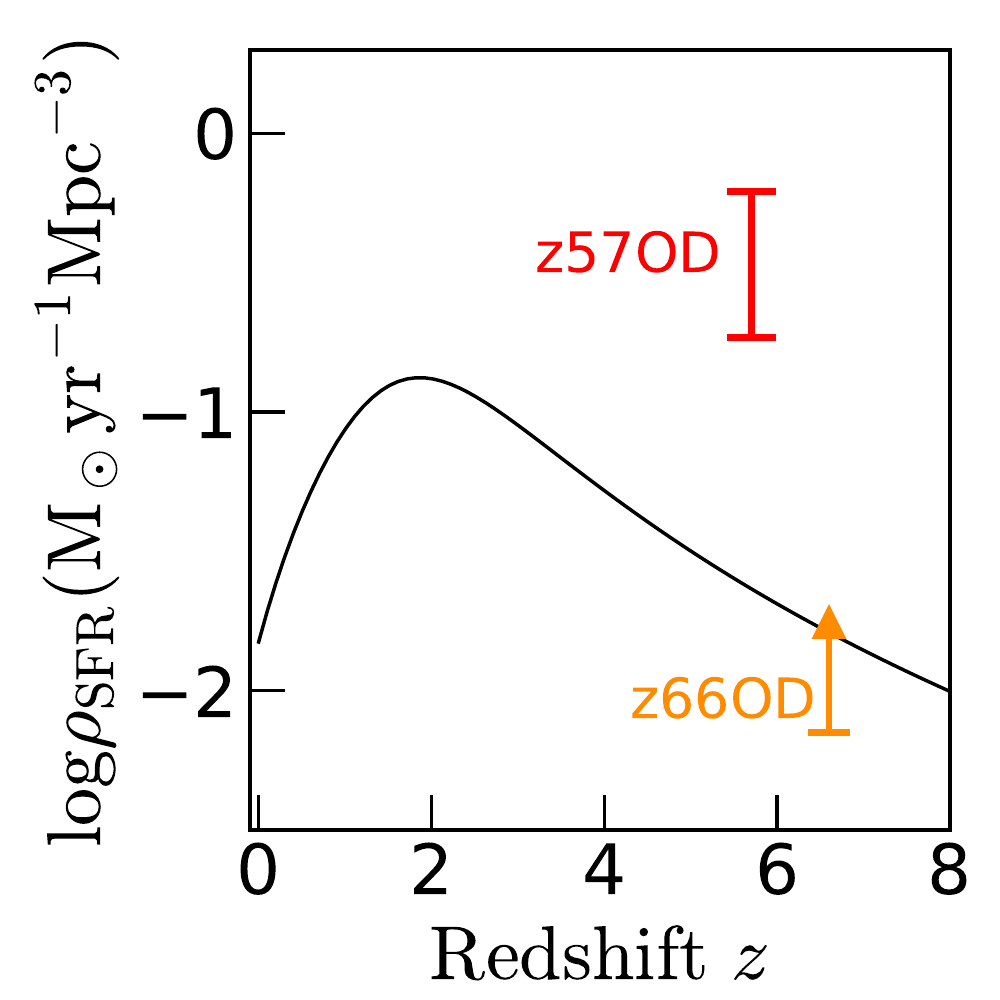}
 \end{center}
   \caption{SFR densities.
   The red bar and orange lower limit are the SFR densities of z57OD and z66OD.
   The red bar is the summation of the observed LAEs and red SMGs with the uncertainty of the fraction of the red SMGs residing at $z=5.7$ ($30-100\%$).
   The orange lower limit only takes account for the observed LAEs.
   Note that we do not include contributions from faint galaxies not detected in our data.
   The black curve is the cosmic average of the SFR density \citep{2014ARA&A..52..415M}.
   The SFR density of z57OD is more than $\sim10$ times higher than the cosmic average.
      \label{fig_cSFR}}
\end{figure}

In Figure \ref{fig_cSFR}, we plot the measured SFR densities as a function of the redshift.
The SFR density in z57OD is $\sim10$ times higher than the cosmic average \citep{2014ARA&A..52..415M}.
We do not obtain a meaningful constraint for z66OD.
These results indicate that star formation is enhanced at least in z57OD.
This active star formation in the overdense region may be explained by high inflow rates in the overdense region.
Recent observational studies reveal that there are tight correlations between the gas accretion rate and star formation rate \citep{2018PASJ...70S..11H,2018arXiv180603299T,2018arXiv180607893B}.
Enhanced star formation of LAEs in the overdense region may be due to high inflow rates in overdensities whose halo is massive.
Indeed the halo masses of z57OD and z66OD are expected to be $4-10\times10^{11}\ \m{M_\odot}$ (see Section \ref{ss_sim}), larger than those of LAEs in normal fields, $1\times10^{11}\ \m{M_\odot}$ \citep{2018PASJ...70S..13O}.

\section{Summary}\label{ss_summary}
We have obtained 3D maps of the universe in the $\sim200\times200\times80$ cMpc$^3$ volumes each at $z=5.7$ and $6.6$ based on the spectroscopic sample of 179 LAEs that accomplishes the $>80$\% completeness down to $\log(L_{\rm Ly\alpha}/[\mathrm{erg\ s^{-1}}])=43.0$, \redc{based on our Keck and Gemini observations and the literature}.
We compare spatial distributions of our LAEs with SMGs, investigate the stellar populations, and compare our LAEs with the numerical simulations.
Our major findings are summarized below.

\begin{enumerate}

\item The 3D maps reveal filamentary large-scale structures extending over $40\ \m{cMpc}$ and two remarkable overdensities made of at least 44 and 12 LAEs at $z=5.692$ (z57OD) and $z=6.585$ (z66OD), respectively.
z66OD is the most distant overdensity spectroscopically confirmed to date \redc{with $>10$ spectroscopically confirmed galaxies}.

\item We have identified similar overdensities to z57OD and z66OD in the simulations regarding the overdensity significance and the velocity dispersion of LAEs.
The halo masses of the overdensities in simulations are $\sim(4-10)\times10^{11}\ \m{M_\odot}$, which will grow to cluster-scale halos ($M_\m{h}\sim10^{14}\ M_\odot$) at the present day, suggesting that z57OD and z66OD are protoclusters.

\item We have selected 44 red 850 $\mu$m-selected SMGs that are SMGs expected to reside at $z\simeq 4-6$ based on their red {\it Herschel} color, and calculated the cross correlation functions between the LAEs and the red SMGs.
We have detected $99.97\%$ cross-correlation signal between $z=5.7$ LAEs and the red SMGs.
This significant correlation suggests that the dust-obscured SMGs are also tracing the same large scale structures as the LAEs, which are typically dust-poor star forming galaxies.

\item Stellar population analyses suggest that LAEs in z57OD and z66OD are actively forming stars with SFRs $\sim5$ times higher than the main sequence at a fixed stellar mass.
Given the significant correlation between the LAEs and the red SMGs at $z=5.7$, the SFR density in z57OD is $10$ times higher than the cosmic average (a.k.a. the Madau-Lilly plot).
\end{enumerate}

\acknowledgments
We thank the anonymous referee for a careful reading and valuable comments that improved the clarity of the paper.
We are grateful to Renyue Cen, Yi-Kuan Chiang, Tadayuki Kodama, and Ken Mawatari for their useful comments and discussions.

The Hyper Suprime-Cam (HSC) collaboration includes the astronomical communities of Japan and Taiwan, and Princeton University.  The HSC instrumentation and software were developed by the National Astronomical Observatory of Japan (NAOJ), the Kavli Institute for the Physics and Mathematics of the Universe (Kavli IPMU), the University of Tokyo, the High Energy Accelerator Research Organization (KEK), the Academia Sinica Institute for Astronomy and Astrophysics in Taiwan (ASIAA), and Princeton University.  Funding was contributed by the FIRST program from Japanese Cabinet Office, the Ministry of Education, Culture, Sports, Science and Technology (MEXT), the Japan Society for the Promotion of Science (JSPS),  Japan Science and Technology Agency  (JST),  the Toray Science  Foundation, NAOJ, Kavli IPMU, KEK, ASIAA,  and Princeton University.


This paper makes use of software developed for the Large Synoptic Survey Telescope. We thank the LSST Project for making their code available as free software at http://dm.lsst.org.

This work is based on observations obtained at the Gemini Observatory processed using the Gemini IRAF package, which is operated by the Association of Universities for Research in Astronomy, Inc., under a cooperative agreement with the NSF on behalf of the Gemini partnership: the National Science Foundation (United States), the National Research Council (Canada), CONICYT (Chile), Ministerio de Ciencia, Tecnolog\'{i}a e Innovaci\'{o}n Productiva (Argentina), and Minist\'{e}rio da Ci\^{e}ncia, Tecnologia e Inova\c{c}\~{a}o (Brazil). 

This work is supported by World Premier International Research Center Initiative (WPI Initiative), MEXT, Japan, and KAKENHI (15H02064, 17H01110, and 17H01114) Grant-in-Aid for Scientific Research (A)
through Japan Society for the Promotion of Science (JSPS).
Y.H. acknowledges support from the Advanced Leading Graduate Course for Photon Science (ALPS) grant and the JSPS through the JSPS Research Fellowship for Young Scientists.
N.K. acknowledges supports from the JSPS grant 15H03645.
I.R.S. acknowledges supports from STFC (ST/P000541/1) and the ERC Advanced Grant DUSTYGAL (321334).
M.I. acknowledges the support from the National Research Foundation of Korea (NRF) grant, No. 2017R1A3A3001362.

\bibliography{ms_himikoOD_v15}

\clearpage

\begin{turnpage}
\tabletypesize{\scriptsize}
\begin{deluxetable*}{cccccccccccc}
\fontsize{8pt}{8pt}\selectfont
\setlength{\tabcolsep}{0.15cm}
\renewcommand{\arraystretch}{0.9}
\tablecaption{An Overview of High Redshift Protoclusters}
\tablehead{\colhead{Name} & \colhead{$z$} & \colhead{$N_\mathrm{spec}$} & \colhead{$\delta$} & \colhead{$\mathrm{Sample}$} & \colhead{$\mathrm{Window size}$} & \colhead{$dz$} & \colhead{$\sigma_\mathrm{V}$}  & \colhead{$M_\mathrm{h}$}  & \colhead{Ref.} \\
\colhead{(1)}& \colhead{(2)}& \colhead{(3)}& \colhead{(4)} &  \colhead{(5)}& \colhead{(6)}& \colhead{(7)}& \colhead{(8)}& \colhead{(9)}& \colhead{(10)}}
\startdata
\multicolumn{10}{c}{Protocluster with $N_\mathrm{spec}\geq10$}\\
z66OD & 6.59 & 12 & $14.3\pm2.1$ & LAE & $\pi\times4.2^2$ & 0.1 & $670\pm200$ & $5.4\times10^{14}$ & This work\\
HSC-z7PCC26 & 6.54 & 14 & $6.8^{+6.1}_{-3.7}$ & LAE & $\pi\times4.2^2$ & 0.1 & $572$ & $8.4\times10^{14}$ & C17,19,Hi18\\
SDF & 6.01 & 10 & $16\pm7$ & LBG & $6\times6$ & $\sim0.05$ & $647\pm124$ & $(2-4)\times10^{14}$ & To12,14\\
z57OD & 5.69 & 44 & $11.5\pm1.6$ & LAE & $\pi\times4.2^2$ & 0.1 & $1280\pm220$ & $4.8\times10^{14}$ & O05,J18,This work\\
SPT2349-56 & 4.31 & 14 & $>1000$ & SMG & $\pi\times0.16^2$ & 0.1 & $408^{+82}_{-56}$ & $1.16\times10^{13}$ & M18\\
TNJ1338-1942 & 4.11 & 37 & $3.7^{+1.0}_{-0.8}$ & LAE/LBG & $7\times7$($\times2$) & 0.049 & $265\pm65$ & $(6-9)\times10^{14}$ & V02,05,07,M04,Z05,Ov08\\
DRC-protocluster & 4.00 & 10 & $\sim5.5-11.0$ & SMG & $0.61\times0.730$ & $\dots$ & $794$ & $(3.2-4.4)\times10^{13}$ & O18\\
PC217.96+32.3 & 3.79 & 65 & $14\pm7$ & LAE & $\pi\times1.2^2$ & 0.035 & $350\pm40$ & $(0.6-1.3)\times10^{15}$ & Lee14,D16,S19\\
D4GD01 & 3.67 & 11 & $\dots$ & LBG & $\pi\times1.8^2$ & $\sim1$ & $352\pm140$ & $\dots$ & To16\\
ClJ0227-0421 & 3.29 & 19 & $10.5\pm2.8$ & Spec & $\pi\times6.2^2$ & 0.09 & $995\pm343$ & $(1.9-3.3)\times10^{14}$ & Lem14\\
TNJ2009-3040 & 3.16 & $>11$ & $0.7^{+0.8}_{-0.6}$ & LAE & $7\times7$ & 0.049 & $515\pm90$ & $\dots$ & V07\\
MRC0316-257 & 3.13 & 31 & $2.3^{+0.5}_{-0.4}$ & LAE & $7\times7$ & 0.049 & $640\pm195$ & $(3-5)\times10^{14}$ & V05,07\\
SSA22FLD & 3.09 & $>15$ & $3.6^{+1.4}_{-1.2}$ & LBG/LAE/SMG & $11.5\times9$ & 0.034 & $\dots$ & $(1.0-1.4)\times10^{15}$ & S98,00,M05,Y12,U17,18\\
MRC0943-242 & 2.92 & 28 & $2.2^{+0.9}_{-0.7}$ & LAE & $7\times7$ & 0.056 & $715\pm105$ & $(4-5)\times10^{14}$ & V07\\
P2Q1 & 2.90 & 12 & $12\pm2$ & Spec & $7\times8$ & 0.016 & $270\pm80$ & $8.1\times10^{14}$ & C14\\
MRC0052-241 & 2.86 & 37 & $2.0^{+0.5}_{-0.4}$ & LAE & $7\times7$ & 0.054 & $980\pm120$ & $(3-4)\times10^{14}$ & V07\\
HS1549 & 2.85 & 26 & $\sim5$ & LBG/SMG & $\dots$ & 0.060 & $\dots$ & $\dots$ & M13,Lac18\\
PCL1002 & 2.45 & 11 & $10$ & Spec/LAE/SMG & $\pi\times2.8^2$ & 0.016 & 426 & $10^{14}-10^{15}$ & D15,Ch15,Ca15\\
HS1700FLD & 2.30 & 19 & $6.9\pm2.1$ & BX/SMG & $8\times8$ & 0.030 & $\dots$ & $1.4\times10^{15}$ & S05,Lac18\\
PKS1138-262 & 2.16 & 15 & $3\pm2$ & LAE/HAE/SMG & $7\times7$ & 0.053 & $900\pm240$ & $(3-4)\times10^{14}$ & K00,04a,04b,P00,02,V07,K13,Z18\\
\hline
\multicolumn{10}{c}{Protocluster with $N_\mathrm{spec}<10$}\\
A2744z8OD & $8.38$ & 1 & $132^{+66}_{-51}$ & LBG & $\pi\times0.1^2$ & $\sim1$ & $\dots$ & $9\times10^{13}$ & I16,L17\\
Borg & $\sim8$ & 0 & $\sim4.5$ & LBG & $2.1\times2.3$ & $\sim1$ & $\dots$ & $>2\times10^{14}$ & Tr12\\
BDF & 7.04 & 3 & $\sim3-4$ & LBG & $\dots$ & $\sim1$ & $\dots$ & $\dots$ & C18\\
HSC-z7PCC4 & 6.58 & 1 & $9.0^{+6.5}_{-4.1}$ & LAE & $\pi\times4.2^2$ & 0.1 & $\dots$ & $\dots$ & Hi18\\
CFHQSJ2329-0301 & 6.43 & 0 & $\sim6$ & LBG & $34\times27$ & $\sim1.0$ & $\dots$ & $\dots$ & U10\\
HSC-z6PCC4 & 5.72 & 4 & $9.7^{+8.5}_{-5.1}$ & LAE & $\pi\times4.2^2$ & 0.1 & $\dots$ & $\dots$ & Hi18\\
HSC-z6PCC5 & 5.69 & 2 & $9.7^{+8.5}_{-5.1}$ & LAE & $\pi\times4.2^2$ & 0.1 & $\dots$ & $\dots$ & Hi18,P18\\
COSMOSAzTEC03 & 5.30 & 4 & $\dots$ & SMG & $1\times1$ & $\dots$ & $\dots$ & $\dots$ & C11\\
TNJ0924-2201 & 5.19 & 6 & $1.5^{+1.6}_{-1.0}$ & LAE/LBG & $7\times7$ & 0.073 & $305\pm110$ & $(4-9)\times10^{14}$ & V04,07,Ov06\\
SDF & 4.86 & 0 & $2.0^{+1.0}_{-2.0}$ & LAE & $10\times10$ & 0.060 & $\dots$ & $>3\times10^{14}$ & S03\\
PClJ1001+0220 & 4.57 & 9 & $3.30\pm0.32$ & Spec & $11\times11$ & 0.01 & $1038\pm178$ & $2.5\times10^{14}$ & Lem18\\
6C0140+326 & 4.41 & 0 & $8\pm5$ & LAE & $10\times10$ & $\sim0.04$ & $\dots$ & $(0.8-2.9)\times10^{14}$ & K11\\
D4UD01 & 3.24 & 5 & $\dots$ & LBG & $\pi\times1.6^2$ & $\sim1$ & $61\pm105$ & $\dots$ & To16\\
D1UD01 & 3.13 & 5 & $\dots$ & LBG & $\pi\times1.6^2$ & $\sim1$ & $235\pm75$ & $\dots$ & To16\\
LABd05 & 2.7 & 0 & $\sim2$ & LAE & $28\times11$ & 0.165 & $\dots$ & $\dots$ & P08\\
USS1558-003 & 2.53 & 0 & $\dots$ & HAE & $7\times4$ & 0.041 & $\dots$ & $\dots$ & H12\\
4C23.56 & 2.48 & 3 & $4.3^{+5.3}_{-2.6}$ & HAE/SMG & $7\times4$ & 0.035 & $\dots$ & $\dots$ & T11,Z18\\
J2143-4423 & 2.38 & 0 & $5.8\pm2.5$ & LAE & $44\times44$ & 0.044 & $\dots$ & $\dots$ & P04\\
4C10.48 & 2.35 & 0 & $11^{+2}_{-2}$ & HAE & $2.5\times2.5$ & 0.046 & $\dots$ & $\dots$ & H11\\
Bo\"oetesJ1430+3522 & 2.3 & 0 & $2.7\pm1.1$ & LAE & $\pi\times5^2$ & 0.0037 & $\dots$ & $1.51\times10^{15}$ & B17
\enddata
\tablecomments{(1) Object name. 
(2) Redshift.
(3) Number of spectroscopically confirmed galaxies.
(4) Galaxy overdensity.
(5) Method of sample selection: (LAE) narrowband LAE, (HAE) narrowband H$\alpha$ emitter, (LBG) Lyman break galaxy, (BX) `BX' galaxy of Adelberger et al. (2005), (SMG) sub-millimeter galaxy, (Spec) spectroscopic survey.
(6) Approximate field size or the size of the structure used to calculate overdensity in units of arcmin$^2$.
(7) Full width redshift uncertainty associated with the $\delta$ quoted.
(8) Velocity dispersion (where available) in units of $\m{km\ s^{-1}}$.
(9) Inferred total halo mass of the overdensity or expected halo mass at $z=0$ in units of M$_\odot$.
(10) Reference (B17:\citealt{2017ApJ...845..172B}, 
C11:\citealt{2011Natur.470..233C}, 
C14:\citealt{2014A&A...570A..16C},
Ca15:\citealt{2015ApJ...808L..33C}, 
Ch15:\citealt{2015ApJ...808...37C},  
C18:\citealt{2018ApJ...863L...3C}, 
C17,19:\citealt{2017MNRAS.469.2646C,2019ApJ...877...51C}, 
D15:\citealt{2015ApJ...802...31D}, 
D16:\citealt{2016ApJ...823...11D}, 
H11:\citealt{2011MNRAS.415.2993H}, 
H12:\citealt{2012ApJ...757...15H}, 
Hi18:\citealt{2018arXiv180100531H}, 
I16:\citealt{2016ApJ...822....5I}, 
J18:\citealt{2018NatAs.tmp..140J}, 
K00,04a,04b:\citealt{2000A&A...358L...1K,2004A&A...428..817K,2004A&A...428..793K}, 
K11:\citealt{2011MNRAS.417.1088K}, 
K13:\citealt{2013MNRAS.434..423K}, 
Lee14:\citealt{2014ApJ...796..126L}, 
Lem14:\citealt{2014A&A...572A..41L}, 
L17:\citealt{2017ApJ...837L..21L}
Lac18:\citealt{2018arXiv180906882L}, 
Lem18:\citealt{2018A&A...615A..77L}
M04:\citealt{2004Natur.427...47M}, 
M05:\citealt{2005ApJ...634L.125M}, 
M13:\citealt{2013ApJ...779...65M}, 
M18:\citealt{2018Natur.556..469M}, 
O05:\citealt{2005ApJ...620L...1O}, 
Ov06,08:\citealt{2006ApJ...637...58O,2008ApJ...673..143O}, 
O18:\citealt{2018ApJ...856...72O}, 
P00,02:\citealt{2000A&A...361L..25P,2002A&A...396..109P}, 
P04:\citealt{2004ApJ...602..545P}, 
P08:\citealt{2008ApJ...678L..77P}, 
P18:\citealt{2018ApJ...861...43P}, 
S98,00,05:\citealt{1998ApJ...492..428S,2000ApJ...532..170S,2005ApJ...626...44S}, 
S03:\citealt{2003ApJ...586L.111S}, 
S19:\citealt{2019arXiv190506337S}, 
T11:\citealt{2011PASJ...63S.415T}, 
To12,14,16:\citealt{2012ApJ...750..137T,2014ApJ...792...15T,2016ApJ...826..114T}, 
Tr12:\citealt{2012ApJ...746...55T}
U10:\citealt{2010ApJ...721.1680U}, 
U17,18:\citealt{2017ApJ...834L..16U,2018PASJ...70...65U}, 
V02,04,05,07:\citealt{2002ApJ...569L..11V,2004A&A...424L..17V,2005A&A...431..793V,2007A&A...461..823V}, 
Y12:\citealt{2012ApJ...751...29Y}, 
Z05:\citealt{2005ApJ...630...68Z}, 
Z18:\citealt{2018MNRAS.479.4577Z})}
\label{tab_PC}
\end{deluxetable*}
\clearpage
\end{turnpage}

\end{document}